\documentclass[aps,prb,twocolumn,floatfix,showpacs]{revtex4}
\usepackage{amsfonts}
\usepackage{amsmath}
\usepackage{amssymb}
\usepackage{graphicx}
\usepackage{color}

\begin{document}

\title{The extended Lawrence-Doniach model: the temperature evolution of the
in-plane magnetic field anisotropy}
\author{M. D. Croitoru$^{1}$}
\author{A. I. Buzdin$^{1,2}$}
\affiliation{$^{1}$Universit\'{e} Bordeaux I, LOMA, UMR 5798, F-33400 Talence, France}
\affiliation{$^2$Institut Universitaire de France, Paris }

\begin{abstract}
Using the quasi-classical formalism we provide the description of the
temperature and field-direction dependence of the in-plane upper critical
field in layered superconductors, taking into account the interlayer
Josephson coupling and the paramagnetic spin splitting. We generalize the
Lawrence-Doniach model for the case of high magnetic fields and show that
the re-entrant superconductivity is naturally described by our formalism
when neglecting the Pauli pair breaking effect. We demonstrate that in
layered superconductors the in-plane anisotropy of the onset of
superconductivity exhibits four different\ temperature regimes: from the
Ginzburg-Landau type in the vicinity of the critical temperature $T_{c0}$
with anisotropies of coherence lengths, up to the FFLO type induced by the
strong interference between the modulation vector and the orbital effect.
Our results are in agreement with the experimental measurements of the
field-angle dependence of the superconducting onset temperature of the
organic compound (TMTSF)$_{2} $ClO$_{4}$.
\end{abstract}

\date{\today }
\author{}
\maketitle


\section{Introduction}

\label{sec0}

Since the discovery of superconductivity in the first layered compound,\cite%
{gamb} there have been found many types of superconductors consisting of
alternating conducting and insulating layers. Examples include the high-$%
T_{c}$ cuprates,\cite{ioffe,clark} layered ruthenates,\cite{berg} the iron
pnictides and oxypnictides,\cite{kami,taka01,pag}, graphite intercalation
compounds,\cite{han,well,emer} crystalline organic metals,\cite%
{buzdin00,sing00,lebe00} the various types of artificial multi-layers,\cite%
{goz,sma,uno} etc. Amongst them, layered organic metals are distinctive for
a number of reasons. Most of them exhibit profound reduced dimensionality
reflected in the very strong charge-transfer anisotropy. The interplay
between electronic correlations and enhanced dimensionality effects leads to
a broad range of physical properties observed in these materials. Moreover,
organic metals are often available in highly clean single crystals that
enables one to perform detailed band-structure measurements and to study
mechanisms of superconductivity in quasi-low-dimensional electronic systems.
Finally one of the most prominent property of organic layered
superconductors is their robustness against high magnetic fields applied
parallel to the conduction layers. Commonly known examples include Bechgaard
salt superconductors (TMTSF)$_{2}$X, where anion X is PF$_{6}$, ClO$_{4}$\
etc. A very large upper critical fields, which exceed the Pauli paramagnetic
limit, for a magnetic field aligned parallel to their conducting layers were
reported.\cite{lee04,oh,yone01} In the compound (TMTSF)$_{2}$PF$_{6}$ $%
H_{c2}=90$~kOe,\cite{lee04} which is more than 4 times larger than $%
H_{P}\simeq 22$~kOe and an enhancement of almost two times over $H_{P}\simeq
27$~kOe is observed in the compound (TMTSF)$_{2}$ClO$_{4}$, $H_{c2}\simeq 50$%
~kOe.\cite{yone01,yone02}

In magnetic field the superconductivity in usual type II superconductors is
suppressed due to the diamagnetic currents and the Pauli pair breaking
effect for singlet pairing. In layered conductors the spatial orbital motion
of electrons is mostly restricted to the conducting planes, when charge
carrier hopping between adjacent layers is small, and the magnetic field
applied precisely parallel to the conducting planes weakly affects the
orbital motion of electrons. Hence the orbital depairing is largely avoided
(there is no magnetic flux inside the 2D Cooper pairs located in planes in
such situation). Moreover, when the interlayer coherence length in a
quasi-1D superconductor is comparable to the interlayer distance the
field-induced quasi-2D (3D) $\rightarrow $ 2D dimensional crossover occurs
in a high magnetic field, restoring the bare critical temperature, $T_{c0}$.%
\cite{gork01,lebe01}

Various theories based on different pairing symmetries predicting the
existence of high-field superconducting state have been proposed previously.
Among them, a phase transition to an inhomogeneous FFLO
(Fulde-Ferell-Larkin-Ovchinnikov) phase for $T<T^{\ast }\simeq 0.56T_{c0}$
or $H>H^{\ast }\simeq 1.06T_{c0}/\mu _{B}$, in which the the singlet
superconducting ground state is characterized by the spatially modulated
order parameter and the spin-polarization. Therefore superconducting state
can be stable beyond the field set by the Pauli paramagnetic limit, $\mu
_{B}H_{P}=\Delta _{0}/\sqrt{2}$, where $\Delta _{0}$ is the superconducting
gap at $T=0$.\cite{lark01,fuld} Conditions for the stabilization of the FFLO
phase are rather stringent,\cite{grun} namely (\textit{i}) the orbital pair
breaking effect should be sufficiently weaker than the Pauli paramagnetic
limit, the Maki parameter $\alpha _{M}\equiv \sqrt{2}H_{c2}/H_{P}\gtrsim 1.8$%
; (\textit{ii}) the system should be in a clean limit.\cite%
{asla,taka,adac,akte03,houz,cui} A growing body of experimental evidence for
the FFLO phase reported from a various measurement techniques supports this
scenario.\cite%
{sing01,sing02,tana,bian,steg,uji01,shin02,lort,agos01,brow01,bergk,agos02,agos03,uji02}
An alternative to the FFLO phase is a triplet pairing state, when the Pauli
spin-splitting destructive mechanism is absent. Within this pairing
symmetry, as was shown by Lebed,\cite{lebe01,lebe06} the superconducting
state is always stable at low temperatures and exhibits a strong re-entrant
behavior in high magnetic field. So far the re-entrant superconducting phase
has not been experimentally identified, at least it is difficult to make
more than a tentative judgement.\cite{lee01,lee06} Nevertheless it can
reveal itself in a number of nontrivial effects in singlet-paired organic
materials in high magnetic fields.\cite{lebe03,lebe04} It was shown that it
can appear in a hidden form and be responsible for an increase of the
superconducting transition temperature in a magnetic field if the orbital
effects of an electron motion are stronger than the Pauli spin-splitting
effects (Paramagnetic intrinsic Meissner effect).\cite{lebe03}

Hitherto there is no experiment which unequivocally answer on the lingering
question concerning the superconducting pairing symmetry in (TMTSF)$_{2}$X
compounds. Previously it was reported that the Knight shift in (TMTSF)$_{2}$%
PF$_{6}$\ conductor does not change at transition temperature supporting the
triplet scenario of pairing.\cite{lee04} However, later experiment with
(TMTSF)$_{2}$ClO$_{4}$ conductor at low-field regime have revealed a clear
change of the Knight shift at the superconducting transition making possible
consideration of the singlet scenario of pairing in such structures.\cite%
{shin02} In the high-field regime the Knight shift is quite weak. On the
other hand, as shown in Ref. \cite{shim06} a small fraction of the triplet
pairing in the singlet paired superconductor strongly enhances the upper
critical field and the triplet component of the order parameter is always
generated in singlet superconductors due to the Pauli paramagnetic
spin-splitting effects.\cite{lebe07,kaba}

In this paper we extend results presented in our previous Letter,\cite%
{vrnkch01} and investigate the in-plane magnetic field-angle dependence of
the onset of superconductivity in layered conductors in the conventional and
the FFLO modulated phases. For this purpose we provide the quasiclassical
description of the anisotropy of the in-plane critical field in layered
superconductors and generalize the Lowerence-Doniach model for the case of
high magnetic fields.

The layout of our paper is as follows. In Sec.\ref{sec1}, we outline our
model based on the quasi-classical formalism for layered superconducting
samples. In Sec. \ref{sec2} we derive the generalized Lowerence-Doniach
equation. In sec. \ref{sec3} we extend this model to the extremely high
magnetic fields. In Sec. \ref{sec4} we focus on the in-plane anisotropy of
the upper critical field for layered superconductors when only orbital
motion is included in the model, and then we investigate the in-plane
anisotropy of $H_{c2}$ when both orbital and paramagnetic depairing are
accounted for. Finally, a short summary is given, where we emphasize the
significance of the obtained results for the interpretation of experiments
with layered superconductors.


\section{General settings}

\label{sec1}

\begin{figure}[tbp]
\resizebox{1.0\columnwidth}{!}{\rotatebox{0}{
\includegraphics{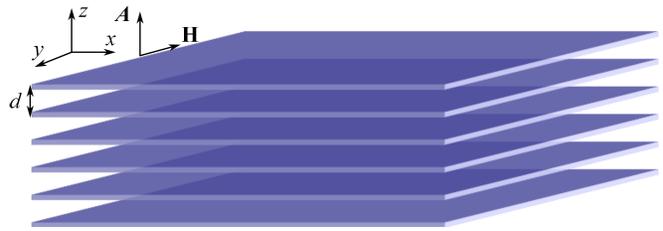}}}
\caption{Scheme of the quasi-2D layered metal.}
\label{fig.01}
\end{figure}

\begin{figure}[tbp]
\resizebox{0.45\columnwidth}{!}{\rotatebox{0}{
\includegraphics{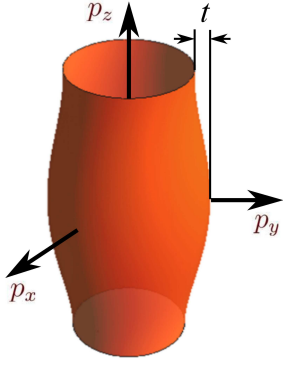}}}
\caption{The Fermi surface of the layered metal in the form of a corrugated
cylinder.}
\label{fig.02}
\end{figure}

We consider a system consisting of layers with good conductivity in \textit{%
xy}-plane stacked along the \textit{z}-axis [see Fig.~\ref{fig.01}]. The
single-electron spectrum is taken as follows%
\begin{equation}
E_{\mathbf{p}}=\frac{p_{x}^{2}}{2m_{x}}+\frac{p_{y}^{2}}{2m_{y}}+\varepsilon
\left( p_{z}\right) ,  \label{001}
\end{equation}%
where $\varepsilon \left( p_{z}\right) =2t\cos \left( p_{z}d\right) $ with $%
d $ - the interlayer distance. We assume that the coupling between layers is
small [see Fig.~\ref{fig.02}], i.e. $t\ll T_{c0}$, but sufficiently large to
make the mean field treatment justified, $T_{c0}^{2}/E_{F}$ $\ll t$.\cite%
{tsu} Here $T_{c0}$ is the critical temperature of the system at $H=0$. In
purely 2D samples, phase fluctuations destroy the long-range order, however
as shown in Ref. \cite{kats01}, even a very small value of hopping leads to
restoration of superconducting order.

We choose the magnetic field to be parallel to the conducting planes and
with a gauge for which the vector potential $\mathbf{A}=\mathbf{H}\times
\mathbf{r}$ [$\mathbf{r}=(x,y,0)$ is a coordinate in \textit{xy}-plane],
i.e. $A_{z}=-xH\sin \alpha +yH\cos \alpha $, where $\alpha $ is the angle
between the applied field, with amplitude $H$, and \textit{x}-axis. Assuming
that the vector potential varies slowly at the interlayer distances (this
assumption means that we neglect the diamagnetic screening currents and take
the magnetic field as uniform and given by the external field, $H$), and
taking into account that the system is near the second-order phase
transition, we can employ the linearized Eilenberger equation for a layered
superconductor in the presence of the parallel magnetic field (in the
momentum representation with respect to the coordinate \textit{z})\cite{kop}:%
\begin{multline}
\left( \Omega _{n}+\widehat{\Pi }\right) f_{\omega }\left( \mathbf{n},%
\mathbf{r},p_{z},k_{z}\right) =\left\{ \Delta (\mathbf{r},k_{z})\frac{{}}{{}}%
\right. \\
+\left. \frac{\left\langle f_{\omega }\left( \mathbf{n},\mathbf{r}%
,p_{z},k_{z}\right) \right\rangle }{2\tau }\right\} \mathrm{sign}(\omega
_{n}).  \label{006}
\end{multline}%
Here
\begin{equation}
\widehat{\Pi }\equiv \frac{\hbar }{2}\mathbf{v}_{F}.\mathbf{\nabla }+2it\sin
(p_{z}d)\sin (\mathbf{Q.r-}\frac{k_{z}}{2}d),
\end{equation}%
where $\mathbf{Q}=(\pi dH/\phi _{0})[-\sin \alpha ,(m_{x}/m_{y})^{1/2}\cos
\alpha ,0]$ with $\phi _{0}=\pi \hbar c/e$, $h=\mu _{B}H$ is the Zeeman
energy, $\mathbf{v}_{F}=v_{F}\mathbf{n}$ is the in-plane Fermi velocity, $%
\tau $ is the impurity scattering time, and $\Omega _{n}\equiv \omega
_{n}-ih+\mathrm{sign}(\omega _{n})/2\tau $. The order parameter is defined
self-consistently as
\begin{equation}
\frac{1}{\lambda }\Delta \left( \mathbf{r},k_{z}\right) =2\pi T\Re
\sum\limits_{\omega >0}\left\langle f_{\omega }\left( \mathbf{n},\mathbf{r}%
,p_{z},k_{z}\right) \right\rangle ,  \label{008}
\end{equation}%
where $\lambda $ is the pairing constant and the brackets denote averaging
over $p_{z}$ and $\mathbf{n}$,%
\begin{equation}
\left\langle ...\right\rangle \equiv \int\limits_{-\frac{\pi }{d}}^{\frac{%
\pi }{d}}\frac{d~dp_{z}}{2\pi }\int\limits_{0}^{2\pi }\frac{d\alpha }{2\pi }%
\left( ...\right) .  \label{009}
\end{equation}%
We assume that the temperature unit is so chosen that the Boltzmann constant
$k_{B}=1$.

Here we considered a layered superconductor in the clean limit, meaning that
the in-plane mean free path is much larger than the corresponding
intra-plane coherence length, $\xi _{0}^{||}=\hbar v_{F}/(2\pi T_{c0})$.
Therefore the linearized Eilenberger equation for the anomalous Green
function $f_{\omega }(\mathbf{n},\mathbf{r},p_{z},k_{z})$ describing layered
superconducting systems acquires the form
\begin{equation}
\left[ \Omega _{n}+\widehat{\Pi }\right] f_{\omega }(\mathbf{n},\mathbf{r}%
,p_{z},k_{z})=\Delta (\mathbf{r},k_{z})  \label{010}
\end{equation}%
with $\Omega _{n}\equiv \omega _{n}-ih~\mathrm{sign}(\omega _{n})$ from now
on.


\section{A layered superconductor in a parallel magnetic field}

\label{sec2}

The upper critical field corresponds to the highest value of $H$, for which
the solution of Eqs. (\ref{008}) and (\ref{010}) exists. To start with, we
consider Eq. (\ref{010}) and write it in the form
\begin{equation}
f_{\omega }(\mathbf{n},\mathbf{r},p_{z},k_{z})=\frac{\Delta (\mathbf{r}%
,k_{z})}{\Omega _{n}}-\frac{1}{\Omega _{n}}\widehat{\Pi }f_{\omega }(\mathbf{%
n},\mathbf{r},p_{z},k_{z})  \label{011}
\end{equation}%
convenient for the subsequent derivation of iterative procedure. Using this
equation we construct the following iterative scheme
\begin{equation}
f_{\omega }^{\left( k+1\right) }\left( \mathbf{n},\mathbf{r}%
,p_{z},k_{z}\right) =\frac{\Delta (\mathbf{r},k_{z})}{\Omega _{n}}-\frac{1}{%
\Omega _{n}}\widehat{\Pi }f_{\omega }^{\left( k\right) }(\mathbf{n},\mathbf{r%
},p_{z},k_{z}).  \label{012}
\end{equation}%
To obtain the convergent iterative scheme we need to require that $\Delta (%
\mathbf{r})\gg \hbar \mathbf{v}_{F}.\bm{\nabla }\Delta (\mathbf{r})/2\pi
T_{c0}$, that implies that characteristic scale of the order parameter
variations should be much larger than $\xi _{0}^{||}=\hbar v_{F}/2\pi T_{c0}$%
. After the completion of the \textit{k}-th iteration we obtain%
\begin{equation}
f_{\omega }^{\left( k+1\right) }\left( \mathbf{n},\mathbf{r}%
,p_{z},k_{z}\right) =\sum\limits_{l=0}^{k}\frac{\left( -1\right) ^{l}}{%
\Omega _{n}^{l+1}}\widehat{\Pi }^{l}\Delta (\mathbf{r},k_{z}).  \label{013}
\end{equation}%
Taking into account the averaging procedure over momentum $p_{z}$, and hence
omitting the terms with even powers of $\sin \left( p_{z}d\right) ,$ then
retaining terms up to the second order in $\hbar \mathbf{v}_{F}.\bm{\nabla }%
\Delta (\mathbf{r})/2\pi T_{c0}$, and making use of the self-consistency
relation Eq. (\ref{008}), we obtain the extended Lowerence-Doniach equation
(MLD equation) in the isotropic case
\begin{multline}
\Delta \left( \mathbf{r},k_{z}\right) \ln \frac{T_{c}}{T_{c0}}=\Delta \left(
\mathbf{r},k_{z}\right) \\
\times \pi T_{c}\sum\limits_{n}\left[ \frac{1}{\omega _{n}}-\frac{1}{\Omega
_{n}}\right] +\widehat{\Pi }_{\mathrm{MLD}}^{h\neq 0}\Delta \left( \mathbf{r}%
,k_{z}\right)  \label{014}
\end{multline}%
where $T_{c0}$ is the critical temperature in the absence of coupling
between adjacent layers, $t$, and of the magnetic field, and%
\begin{multline}
\widehat{\Pi }_{\mathrm{MLD}}^{h\neq 0}\equiv \pi T_{c}\sum\limits_{n}\frac{%
\hbar ^{2}v_{F}^{2}}{8\Omega _{n}^{3}}\nabla ^{2}-\frac{t^{2}}{\Omega
_{n}^{3}}\left[ 1-\cos \left( 2\mathbf{Q.r-}k_{z}d\right) \right] \\
+\frac{\hbar ^{2}\left( v_{F}Q\right) ^{2}}{8}\frac{t^{2}}{\Omega _{n}^{5}}%
\left[ 1-7\cos \left( 2\mathbf{Q.r-}k_{z}d\right) \right]  \label{014a}
\end{multline}%
The anisotropic case one can obtain simply by the following substitutions $%
\hbar ^{2}v_{F}^{2}\nabla ^{2}\Delta \left( \mathbf{r}\right) \rightarrow
2\varepsilon \left( \mathbf{\nabla }\right) \Delta \left( \mathbf{r}\right)
\equiv 2\hbar ^{2}\left\{ \left\langle v_{Fx}^{2}\right\rangle \partial
_{x}^{2}+\left\langle v_{Fy}^{2}\right\rangle \partial _{y}^{2}\right\}
\Delta \left( \mathbf{r}\right) $ and $\hbar ^{2}\left( v_{F}Q\right)
^{2}\rightarrow 2\varepsilon \left( \mathbf{Q}\right) $, where
\begin{equation}
\varepsilon \left( \mathbf{Q}\right) \equiv \hbar ^{2}\left\{ \left\langle
v_{Fx}^{2}\right\rangle Q_{x}^{2}+\left\langle v_{Fy}^{2}\right\rangle
Q_{y}^{2}\right\} .
\end{equation}%
Introducing the temperature $T_{cP}$, as the superconducting onset
temperature in the pure Pauli limit determined by the expression
\begin{equation}
\ln \frac{T_{c0}}{T_{cP}}=\pi T_{cP}\sum\limits_{n}\left[ \frac{1}{\omega
_{n}}-\frac{1}{\Omega _{n}}\right] ,  \label{015a}
\end{equation}%
and the use of the identities $2\pi T\sum\nolimits_{n=0}^{\infty }\Omega
_{n}^{-3}=-\Phi ^{\left( 2\right) }\left( h\right) /8\pi ^{2}T^{2}$ and $%
2\pi T\sum\nolimits_{n=0}^{\infty }\Omega _{n}^{-5}=-\Phi ^{\left( 4\right)
}\left( h\right) /384\pi ^{2}T^{2}$ gives rise to (for details see Appendix %
\ref{sec:appendix_A})
\begin{multline}
\Delta \left( \mathbf{r},k_{z}\right) P=-\frac{\Phi ^{\left( 2\right)
}\left( h\right) }{8\pi ^{2}T_{cP}^{2}} \\
\times \left\{ \frac{\varepsilon \left( \mathbf{\nabla }\right) }{4}-t^{2}%
\left[ 1-\cos \left( 2\mathbf{Q.r-}k_{z}d\right) \right] \right\} \Delta
\left( \mathbf{r},k_{z}\right) \\
-\frac{\Phi ^{\left( 4\right) }\left( h\right) t^{2}}{384\pi ^{4}T_{cP}^{4}}%
\frac{\varepsilon \left( \mathbf{Q}\right) }{4}\left[ 1-7\cos \left( 2%
\mathbf{Q.r-}k_{z}d\right) \right] \Delta \left( \mathbf{r},k_{z}\right) ,
\label{015d}
\end{multline}%
where $P=\left( T_{c}-T_{cP}\right) /AT_{c}$, $\Phi ^{\left( k\right)
}\left( h\right) \equiv \left[ \psi ^{\left( k\right) }\left( 1/2+ih\right)
+\psi ^{\left( k\right) }\left( 1/2-ih\right) \right] /2$ with $\psi
^{\left( k\right) }\left( z\right) =d^{k}\psi \left( z\right) /dz^{k}$ and $%
\psi \left( z\right) $\ is the digamma function. If we can neglect the
Zeeman effect, $h=0$, than $\Omega _{n}\rightarrow \omega _{n}$, and making
use of the identities $2\pi T\sum\nolimits_{n=0}^{\infty }\omega
_{n}^{-3}=7\zeta \left( 3\right) /4\pi ^{2}T^{2}$ and $2\pi
T\sum\nolimits_{n=0}^{\infty }\omega _{n}^{-5}=31\zeta \left( 5\right)
/16\pi ^{2}T^{2}$ reduces Eq. (\ref{015d}) to
\begin{multline}
\Delta \left( \mathbf{r},k_{z}\right) \ln \frac{T_{c}}{T_{c0}}=\frac{7\zeta
\left( 3\right) }{4\pi ^{2}T_{c0}^{2}} \\
\times \left\{ \frac{\varepsilon \left( \mathbf{\nabla }\right) }{4}-t^{2}%
\left[ 1-\cos \left( 2\mathbf{Q.r-}k_{z}d\right) \right] \right\} \Delta
\left( \mathbf{r},k_{z}\right) \\
+\frac{31\zeta \left( 5\right) t^{2}}{16\pi ^{4}T_{c0}^{4}}\frac{\varepsilon
\left( \mathbf{Q}\right) }{4}\left[ 1-7\cos \left( 2\mathbf{Q.r-}%
k_{z}d\right) \right] \Delta \left( \mathbf{r},k_{z}\right) ,  \label{015}
\end{multline}%
where $T_{c0}$ is the superconducting critical temperature in the absence of
coupling between adjacent layers, $t$, and in the absence of the magnetic
field, described by the vector $Q$. As it is seen the MLD equation contains
the term, proportional to $\left( v_{F}Q\right) ^{2}t^{2}$, which is absent
in the standard Lowerence-Doniach equation. As it will be seen later this
term represents unusual orbital contribution responsible for the re-entrant
superconducting phase at high magnetic fields.\cite{lebe03} Let us consider
several limiting cases.


\subsection{Regime $H\ll \frac{t}{\protect\pi \hbar dv_{F}}\protect\phi _{0}$%
}

First, let us consider the case of a small magnetic field. When $\hbar
v_{F}Q $ $\ll T_{c0}$, we can retain only terms up to the second order in $%
\left( \hbar v_{F}Q\right) /T_{c0}$ or/and $t/T_{c0}$. Then after neglecting
the last term in the MLD, because it is much smaller than other terms, Eq. (%
\ref{014}) reduces to the standard Lowerence-Doniach equation
\begin{multline}
\Delta \left( \mathbf{r},k_{z}\right) P=\pi T_{c}\sum\limits_{n}\frac{%
\varepsilon \left( \mathbf{\nabla }\right) }{4\Omega _{n}^{3}}\Delta \left(
\mathbf{r},k_{z}\right) \\
-\frac{t^{2}}{\Omega _{n}^{3}}2\sin ^{2}\left( \mathbf{Q.r}-\frac{k_{z}}{2}%
d\right) \Delta \left( \mathbf{r},k_{z}\right) .  \label{017}
\end{multline}%
In the continuous limit, $d\rightarrow 0$, $d\ll \xi _{0}^{\perp }\left(
T\right) $ with $\xi _{0}^{\perp }$ - the inter-plane coherence length, Eq. (%
\ref{017}) transforms into the Ginzburg-Landau equation for an anisotropic
superconductor. If the order parameter is homogeneous along the \textit{z}%
-axis we can set $k_{z}=0$. If $Qr\mathbf{\sim }Ql\ll 1$, or $H\ll \frac{t}{%
\pi \hbar dv_{F}}\phi _{0}$, where $l=\sqrt{\hbar /m\widetilde{\omega }_{H}}$
is the characteristic magnetic length with the characteristic magnetic
frequency $\widetilde{\omega }_{H}$ defined as $\widetilde{\omega }_{H}=%
\sqrt{\frac{2\gamma _{z}}{m_{x}}T_{c0}}\frac{2\pi }{\phi _{0}}H$, Eq. (\ref%
{017}) can be further simplified
\begin{equation}
P\Delta \left( \mathbf{r}\right) -\left[ \gamma _{x}\partial _{x}^{2}+\gamma
_{y}\partial _{y}^{2}-\gamma _{z}\left( \frac{2\mathbf{Qr}}{d}\right) ^{2}%
\right] \Delta \left( \mathbf{r}\right) =0,  \label{019}
\end{equation}%
where $\alpha =\left( T_{c}-T_{c0}\right) /T_{c0}$, $\gamma _{x,y}=-\hbar
^{2}\Phi ^{\left( 2\right) }\left( h\right) \left\langle
v_{Fx,y}^{2}\right\rangle /32\pi ^{2}T_{cP}^{2}$, $\gamma
_{z}=d^{2}t^{2}\Phi ^{\left( 2\right) }\left( h\right) /16\pi ^{2}T_{cP}^{2}$%
. If $h=0$ we may write $\gamma _{x,y}=\beta \hbar ^{2}\left\langle
v_{Fx,y}^{2}\right\rangle /2T_{c0}^{2}=\beta \hbar ^{2}v_{F}^{2}/4T_{c0}^{2}$%
, $\gamma _{z}=\beta d^{2}t^{2}/T_{c0}^{2}$, where $\beta =7\zeta \left(
3\right) /8\pi ^{2}$. The cyclotron frequency is $\widetilde{\omega }_{H}=%
\sqrt{\frac{\gamma _{z}}{\gamma _{x}}}\frac{\hbar }{m_{x}}\frac{2\pi }{\phi
_{0}}H$, or using the relation $\gamma _{x}/\gamma _{z}=\left\langle
v_{Fx,y}^{2}\right\rangle /2d^{2}t^{2}=v_{F}^{2}/4d^{2}t^{2}$, is $%
\widetilde{\omega }_{H}=\frac{2dt}{m_{x}v_{F}}\frac{2\pi }{\phi _{0}}H$.
After performing scaling of the variable $y^{\prime }=\sqrt{m_{y}/m_{x}}y$,
the anisotropic model with effective masses can be reduced to the isotropic
one in the renormalized magnetic field $H\rightarrow H\sqrt{\sin ^{2}\left(
\vartheta \right) +\frac{m_{x}}{m_{y}}\cos ^{2}\left( \vartheta \right) }$,%
\cite{brison} where $\hbar ^{2}/2m_{x,y}=\gamma _{x,y}T_{c0}$. Finally, the
angle-resolved highest magnetic field, at which superconductivity can
nucleate in a sample is given by
\begin{equation}
\left. H_{c2}\left( \vartheta ,T\right) \right\vert _{\varkappa _{I}}=\frac{%
\left. H_{c2}\left( \frac{\pi }{2}\right) \right\vert _{\varkappa _{I}}}{%
\sqrt{\sin ^{2}\left( \vartheta \right) +\frac{m_{x}}{m_{y}}\cos ^{2}\left(
\vartheta \right) }}.  \label{020}
\end{equation}%
Here for the negligible Zeeman effect, which breaks apart the paired
electrons if they are in a spin-singlet state, $h=0$,%
\begin{equation}
\left. H_{c2}^{h=0}\left( \frac{\pi }{2}\right) \right\vert _{\varkappa
_{I}}=\frac{m_{x}}{\hbar ^{2}}\frac{\hbar v_{F}}{d}\frac{T_{c0}}{t}\frac{%
\phi _{0}}{2\pi }\left( 1-\frac{T_{c}}{T_{c0}}\right) ,  \label{021}
\end{equation}%
while for $h\neq 0$%
\begin{equation}
\left. H_{c2}^{h\neq 0}\left( \frac{\pi }{2}\right) \right\vert _{\varkappa
_{I}}=\frac{8\pi T_{c0}}{A\hbar dt}\sqrt{\frac{m_{x}T_{c0}}{2\Phi ^{\left(
2\right) }\left( h\right) }}\frac{\phi _{0}}{2\pi }\left( 1-\frac{T_{c}}{%
T_{cP}}\right) ,
\end{equation}%
where $\varkappa _{I}:H\ll \frac{t}{\pi \hbar dv_{F}}\phi _{0}$.

\begin{figure}[tbp]
\resizebox{0.85\columnwidth}{!}{\rotatebox{0}{
\includegraphics{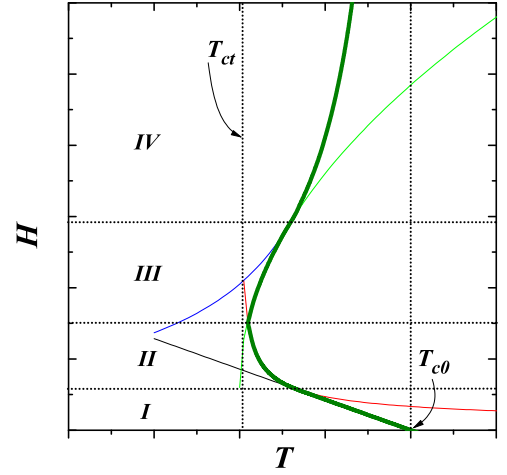}}}
\caption{Scheme of the $H-T$ phase diagram for layered superconductors, when
external magnetic field, $H$, is applied parallel to the layers, $t\ll
T_{c0} $, and the paramagnetic effects are vanished.}
\label{fig.03}
\end{figure}


\subsection{The crossover regime: $\frac{\protect\phi _{0}}{\protect\pi %
\hbar dv_{F}}t\ll H\ll \frac{\protect\phi _{0}}{\protect\pi \hbar dv_{F}}%
T_{c0}$}

To study the anisotropy of the upper critical field, when its amplitude is
in the range, $t\ll \hbar v_{F}Q\ll T_{c0}$, or $\frac{\phi _{0}}{\pi \hbar
dv_{F}}t\ll H\ll \frac{\phi _{0}}{\pi \hbar dv_{F}}T_{c0}$, we employ the
extended Lowerence-Doniach equation Eq. (\ref{014}) and choose the solution
in the form%
\begin{equation}
\Delta \left( \mathbf{r}\right) =\Delta _{0}+\Delta _{2}\cos \left( 2\mathbf{%
Q.r}\right) .  \label{030}
\end{equation}%
By substitution it in Eq. (\ref{014}), we obtain the following system of
coupled equations%
\begin{multline}
\Delta _{0}P=\pi T_{cP}\sum\limits_{n}\left[ -\frac{1}{\Omega _{n}^{3}}+%
\frac{\varepsilon \left( \mathbf{Q}\right) }{4\Omega _{n}^{5}}\right]
t^{2}\Delta _{0} \\
+\pi T_{cP}\sum\limits_{n}\left[ \frac{1}{2\Omega _{n}^{3}}-7\frac{%
\varepsilon \left( \mathbf{Q}\right) }{8\Omega _{n}^{5}}\right] t^{2}\Delta
_{2},  \label{031}
\end{multline}%
and%
\begin{multline}
\Delta _{2}P=-\pi T_{cP}\sum\limits_{n}\left\{ \frac{\varepsilon \left(
\mathbf{Q}\right) }{\Omega _{n}^{3}}\Delta _{2}\right. \\
+\left. \left( \Delta _{0}-\frac{\Delta _{2}}{2}\right) \frac{t^{2}}{\Omega
_{n}^{3}}-\left( 7\Delta _{0}+\frac{5\Delta _{2}}{2}\right) \frac{%
\varepsilon \left( \mathbf{Q}\right) t^{2}}{8\Omega _{n}^{5}}\right\} .
\label{032}
\end{multline}%
In the situation $\left\vert P\right\vert \ll \left\vert \Phi ^{\left(
2\right) }\left( h\right) \right\vert \varepsilon \left( \mathbf{Q}\right)
/8\pi ^{2}T_{cP}^{2}$, when taking into account that $\Delta _{0}\gg \Delta
_{2}$, from Eq. (\ref{032}) we can obtain $\Delta _{2}=t^{2}\Delta
_{0}/\varepsilon \left( \mathbf{Q}\right) .$ Substituting it into Eq. (\ref%
{031}) and retaining only terms up to the second order in ($t/T_{c0}$) leads
to
\begin{equation}
P=\pi T_{cP}\sum\limits_{n}\left[ -\frac{1}{\Omega _{n}^{3}}+\frac{%
\varepsilon \left( \mathbf{Q}\right) }{4\Omega _{n}^{5}}+\frac{1}{2\Omega
_{n}^{3}}\frac{t^{2}}{\varepsilon \left( \mathbf{Q}\right) }\right] t^{2}
\label{033}
\end{equation}%
or%
\begin{equation}
P=\frac{\Phi ^{\left( 2\right) }\left( h\right) }{8\pi ^{2}T_{cP}^{2}}%
t^{2}-\varepsilon \left( \mathbf{Q}\right) t^{2}\frac{\Phi ^{\left( 4\right)
}\left( h\right) }{1536\pi ^{2}T_{cP}^{2}}-\frac{t^{4}}{\varepsilon \left(
\mathbf{Q}\right) }\frac{\Phi ^{\left( 2\right) }\left( h\right) }{16\pi
^{2}T_{cP}^{2}}
\end{equation}%
If the Zeeman effect of the applied field is absent, $h=0$, we have to make
the following substitution, $\Omega _{n}\rightarrow \omega _{n}$ and $%
T_{cP}\rightarrow T_{c0}$. After introducing the temperature, $T_{ct}$,
accounting for the coupling between adjacent layers via expression $\ln
\left( T_{c0}/T_{ct}\right) =t^{2}\pi T_{ct}\sum\nolimits_{n}\omega
_{n}^{-3} $, Eq. (\ref{033}) acquires the form%
\begin{equation}
\ln \frac{T_{c}}{T_{ct}}=t^{2}\pi T_{c0}\sum\limits_{n}\frac{\varepsilon
\left( \mathbf{Q}\right) }{4\omega _{n}^{5}}+\frac{t^{4}}{\varepsilon \left(
\mathbf{Q}\right) }\pi T_{c0}\sum\limits_{n}\frac{1}{2\omega _{n}^{3}},
\label{035}
\end{equation}%
where%
\begin{equation}
\varepsilon \left( \mathbf{Q}\right) \equiv \frac{\hbar ^{2}v_{F}^{2}}{2}%
\frac{\pi ^{2}d^{2}H^{2}}{\phi _{0}^{2}}\left[ \sin ^{2}\left( \vartheta
\right) +\frac{m_{x}}{m_{y}}\cos ^{2}\left( \vartheta \right) \right] ,
\end{equation}%
or using the definition $\left\langle v_{Fx,y}^{2}\right\rangle =\frac{T_{c0}%
}{\beta m_{x,y}}$,

\begin{equation}
\varepsilon \left( \mathbf{Q}\right) \equiv \frac{\hbar ^{2}}{m_{x}}\frac{%
T_{c0}}{\beta }\frac{\pi ^{2}d^{2}H^{2}}{\phi _{0}^{2}}\left[ \sin
^{2}\left( \vartheta \right) +\frac{m_{x}}{m_{y}}\cos ^{2}\left( \vartheta
\right) \right] .
\end{equation}%
Eq. (\ref{035}) is the transcendental equation to determine $H_{c2}\left(
\vartheta ,T\right) $ for a layered system with interlayer coupling $t$. Let
us consider two limiting situations.

\subsubsection{Lowerence-Doniach regime $t\protect\phi _{0}/\protect\pi %
\hbar dv_{F}\ll H\ll \protect\sqrt{tT_{c0}}\protect\phi _{0}/\protect\pi %
\hbar dv_{F}$}

If the amplitude of the external magnetic field satisfies the condition $%
t\ll \hbar v_{F}Q\ll \sqrt{tT_{c0}}$, or $t\phi _{0}/\pi \hbar dv_{F}\ll
H\ll \sqrt{tT_{c0}}\phi _{0}/\pi \hbar dv_{F}$, we can neglect the first
term in Eq. (\ref{035}) and obtain Eq. (\ref{020}), as the expression for
the upper critical field with
\begin{equation}
\left. H_{c2}^{h=0}\left( \frac{\pi }{2}\right) \right\vert _{\varkappa
_{II}}=\frac{7\zeta \left( 3\right) t^{2}}{8\pi ^{3}\hbar dT_{c0}}\sqrt{%
\frac{2m_{x}}{T_{c}-T_{ct}}}\phi _{0},
\end{equation}%
where $\varkappa _{II}:t\phi _{0}/\pi \hbar dv_{F}\ll H\ll \sqrt{tT_{c0}}%
\phi _{0}/\pi \hbar dv_{F}$, when the Zeeman effect is negligible.

\begin{figure}[tbp]
\resizebox{0.85\columnwidth}{!}{\rotatebox{0}{
\includegraphics{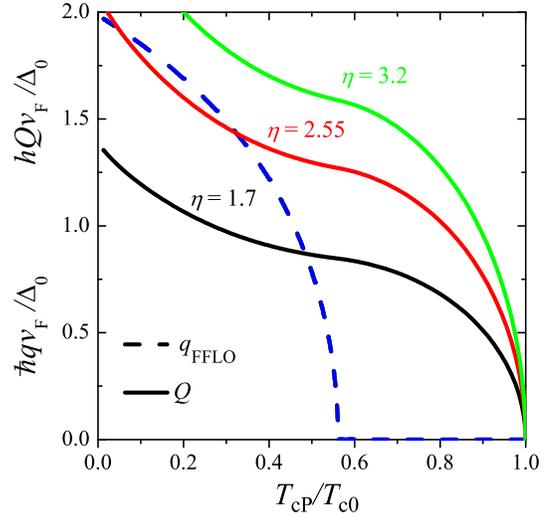}}}
\caption{$\hbar qv_{F}/k_{B}T_{c0}$ and $\hbar Qv_{F}/k_{B}T_{c0}$ versus $%
T_{cP}/T_{c0}$ for several values of $\protect\eta $.}
\label{fig.04}
\end{figure}

\subsubsection{Regime $\frac{\protect\phi _{0}}{\protect\pi \hbar dv_{F}}%
\protect\sqrt{tT_{c0}}\ll H\ll \frac{\protect\phi _{0}}{\protect\pi \hbar
dv_{F}}T_{c0}$}

If the field is such that $\sqrt{tT_{c0}}\ll \hbar v_{F}Q\ll T_{c0}$, or $%
\frac{\phi _{0}}{\pi \hbar dv_{F}}\sqrt{tT_{c0}}\ll H\ll \frac{\phi _{0}}{%
\pi \hbar dv_{F}}T_{c0}$ the expression for the upper critical field, $%
H_{c2}^{{}}\left( \vartheta ,T\right) $, can be obtained from Eq. (\ref{035}%
) by neglecting the second term. Then again we obtain Eq. (\ref{020}), as
the expression for the upper critical field with
\begin{equation}
\left. H_{c2}^{h=0}\left( \frac{\pi }{2}\right) \right\vert _{\varkappa
_{III}}=\sqrt{\frac{28}{31}\frac{\zeta \left( 3\right) }{\zeta \left(
5\right) }}\frac{T_{c0}}{dt}\sqrt{\frac{2m_{x}\left( T_{c}-T_{ct}\right) }{%
\hbar ^{2}}}\phi _{0}.
\end{equation}%
This regime describes the beginning of the reentrant superconductivity
regime.\cite{lebe01,lebe06}


\section{General case for $H\gg \frac{t}{\protect\pi \hbar dv_{F}}\protect%
\phi _{0}$}

\label{sec3}

To study the anisotropy of the upper critical field, when its amplitude
satisfies $H\gg \frac{t}{\pi \hbar dv_{F}}\phi _{0}$ we need to reconsider
the solution of the Eilenberger equation, Eq. (\ref{010}). Since the
magnetic field induced potential has the form $V\left( \mathbf{r}\right)
=t\sin \left( p_{z}d\right) \left[ e^{i\mathbf{Q.r}}-e^{-i\mathbf{Q.r}}%
\right] $ $=2it\sin \left( p_{z}d\right) \sin \left( \mathbf{Q.r}\right) $,
i.e. it is periodic in real space, the solution of Eq. (\ref{010}) can be
written without any loss of generality as\cite{vrnkch01}
\begin{equation}
f_{\omega }\left( \mathbf{n}_{\mathbf{p}},\mathbf{r},p_{z}\right) =e^{i%
\mathbf{q.r}}\sum\limits_{m}e^{im\mathbf{Q.r}}f_{m}\left( \omega _{n},%
\mathbf{n}_{\mathbf{p}},p_{z}\right) ,  \label{050}
\end{equation}%
where we took into account the possibility of the FFLO phase formation in
this field regime. Because of the form for $f_{\omega }\left( \mathbf{n}_{%
\mathbf{p}},\mathbf{r},p_{z}\right) $ of Eq. (\ref{050}) one can write $%
\Delta \left( \mathbf{r}\right) $ as%
\begin{equation}
\Delta \left( \mathbf{r}\right) =e^{i\mathbf{q.r}}\sum\limits_{m}e^{i2m%
\mathbf{Q.r}}\Delta _{2m}.  \label{052}
\end{equation}%
From symmetry considerations it follows that $\Delta _{-2m}=\Delta _{2m}$.
Substituting Eqs. (\ref{050}) and (\ref{052}) back into Eq. (\ref{010}) one
gets\cite{vrnkch02}
\begin{align}
L_{n}\left( \mathbf{q}\right) f_{0}+\widetilde{t}f_{-1}-\widetilde{t}f_{1}&
=\Delta _{0},  \label{060} \\
L_{n}\left( \mathbf{q}\pm \mathbf{Q}\right) f_{\pm 1}\mathbf{\pm }\widetilde{%
t}f_{0}\mp \widetilde{t}f_{\mathbf{\pm }2}& =0,  \label{061} \\
L_{n}\left( \mathbf{q}\pm 2\mathbf{Q}\right) f_{\pm 2}\pm \widetilde{t}%
f_{\pm 1}\mp \widetilde{t}f_{\mathbf{\pm }3}& =\Delta _{\mathbf{\pm }2},
\label{062} \\
L_{n}\left( \mathbf{q}\pm 3\mathbf{Q}\right) f_{\pm 3}\pm \widetilde{t}%
f_{\pm 2}& =0,  \label{063}
\end{align}%
where $f_{m}\equiv f_{m}\left( \omega _{n},\mathbf{n},p_{z}\right) $, $%
L_{n}\left( \mathbf{s}\right) =\Omega _{n}+i\hbar \mathbf{v}_{F}\mathbf{s}/2$
and $\widetilde{t}=t\sin \left( p_{z}d\right) $. Here we took into account
that $\Delta _{\pm \left( 2m+1\right) }=0$. When deriving this set of
coupled equations we accounted for $t\ll \hbar v_{F}Q$, or $\frac{\phi _{0}}{%
\pi \hbar dv_{F}}t\ll H$. This limit allowed us to retain only $\Delta _{0}$
and $\Delta _{\mathbf{\pm }2}$, or $f_{0}$, $f_{\pm 1}$, $f_{\pm 2}$
harmonics, because we adopt a second-order approximation in the small
parameter $t/T_{c0}$ to the solution of Eq. (\ref{010}), $t\ll T_{c0}$).
Actually, if the applied field is such that $T_{c0}\lesssim \hbar v_{F}Q$,
then it would be sufficient to retain only $\Delta _{0}$, or $f_{0}$, $%
f_{\pm 1}$ harmonics.

\begin{figure*}[tbp]
\resizebox{0.66\columnwidth}{!}{\rotatebox{0}{
\includegraphics{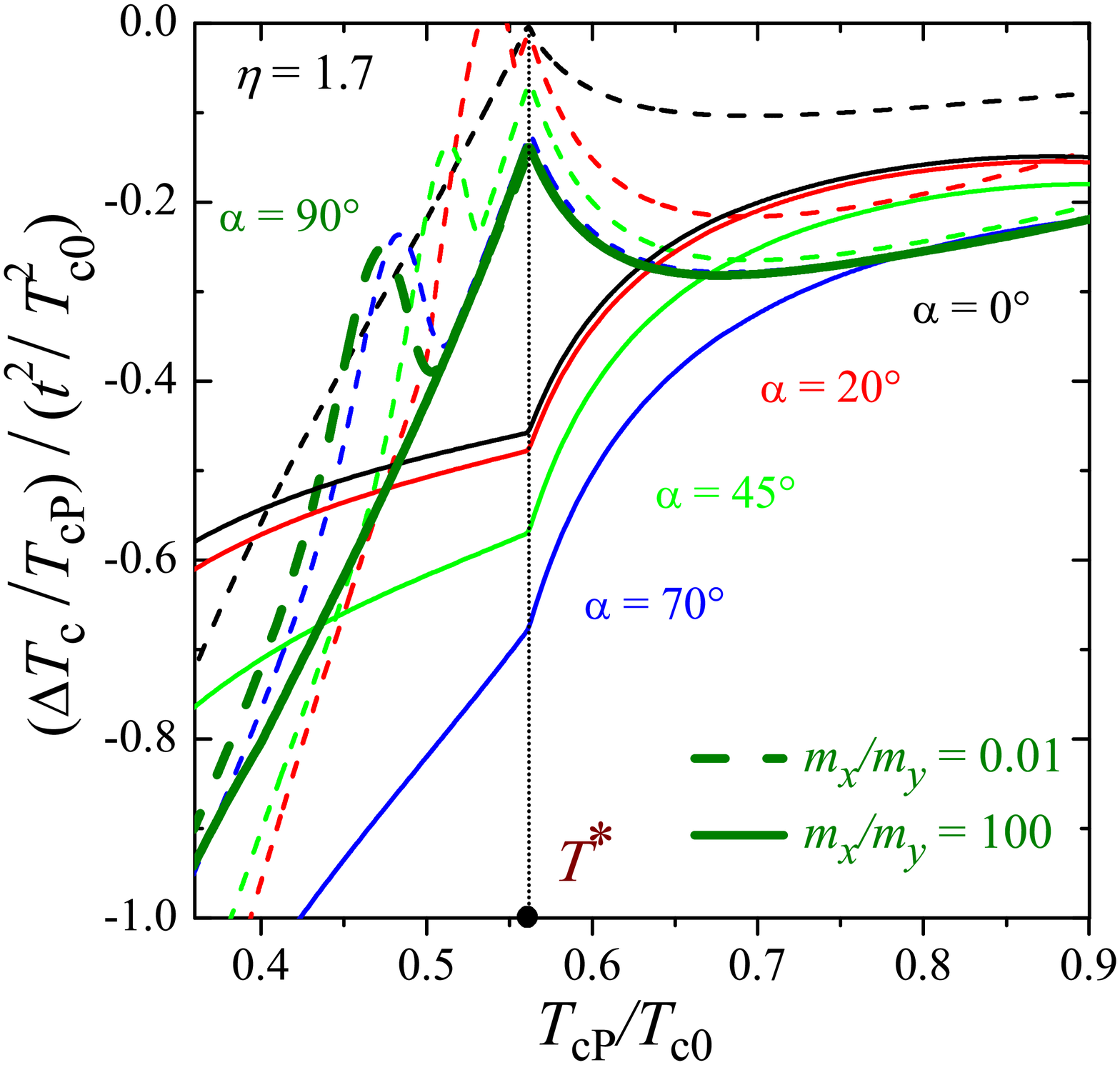}}}
\resizebox{0.66\columnwidth}{!}{\rotatebox{0}{
\includegraphics{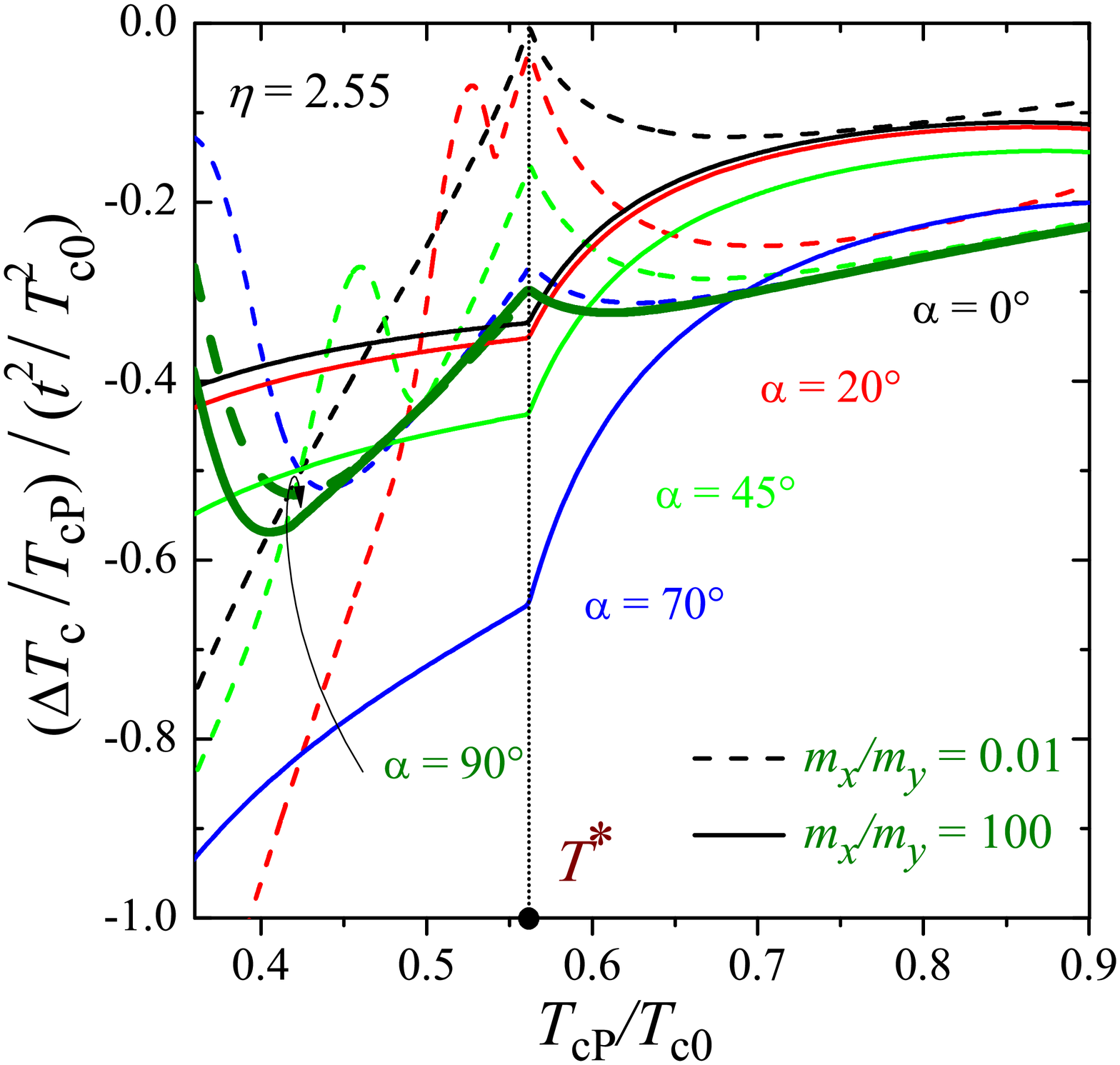}}}
\resizebox{0.66\columnwidth}{!}{\rotatebox{0}{
\includegraphics{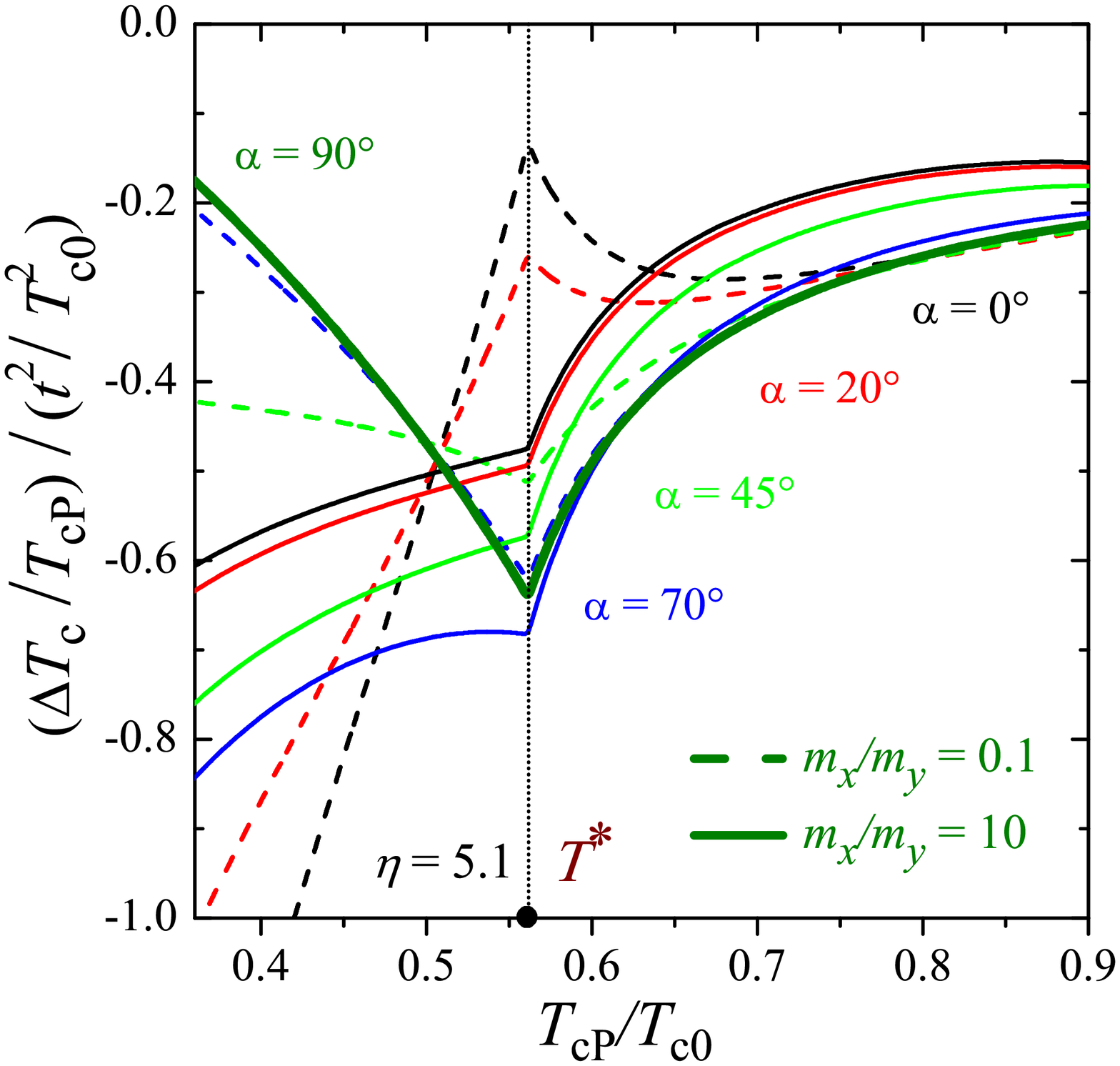}}}
\caption{Contribution of the orbital effect as a function of $T_{cP}/T_{c0}$
for several angles $\protect\alpha $ between $\mathbf{H}$ and $x$-axis, for $%
\protect\eta =1.7$ (left panel), for $\protect\eta =2.5$ (middle panel) ,
and for $\protect\eta =5.1$ (right panel). Solid lines are the results
obtained for $m_{x}/m_{y}=100$ and $m_{x}/m_{y}=10$; Dashed lines are for $%
m_{x}/m_{y}=0.01$, $m_{x}/m_{y}=0.1$. }
\label{fig.05}
\end{figure*}

Making use of the self-consistency relation the solution of the system of
coupled equations (\ref{060} - \ref{063}) can be given in the form (for
details see Appendix \ref{sec:appendix_B})

\begin{align}
\Delta _{0}\left[ P+t^{2}a\right] & =t^{2}\sum\limits_{\pm }c_{\pm }\Delta _{%
\mathbf{\pm }2},  \label{080} \\
\Delta _{\mathbf{+}2}\left[ P+t^{2}b_{+}+\delta _{+}\right] &
=t^{2}c_{+}\Delta _{0},  \label{081} \\
\Delta _{\mathbf{-}2}\left[ P+t^{2}b_{-}+\delta _{-}\right] &
=t^{2}c_{-}\Delta _{0}  \label{082}
\end{align}%
where the following notations are introduced:%
\begin{align}
a& =\pi T\sum\limits_{n,\xi =\pm }\left. T_{n}\left( \mathbf{q},\mathbf{q}%
,\xi \mathbf{Q}\right) \right\vert _{T=T_{cP}},  \label{182} \\
b_{\pm }& =\pi T\sum\limits_{n,\xi =\pm }\left. T_{n}\left( \mathbf{q}\pm 2%
\mathbf{Q},\mathbf{q}\pm 2\mathbf{Q},\mathbf{q}\pm 2\mathbf{Q+}\xi \mathbf{Q}%
\right) \right\vert _{T=T_{cP}}, \\
c_{\pm }& =\pi T\sum\limits_{n}\left. T_{n}\left( \mathbf{q},\mathbf{q}\pm
\mathbf{Q},\mathbf{q}\pm 2\mathbf{Q}\right) \right\vert _{T=T_{cP}}
\label{083} \\
\delta _{\pm }& =\pi T\sum\limits_{n}\left. \frac{1}{L_{n}\left( \mathbf{q}%
\right) }-\frac{1}{L_{n}\left( \mathbf{q}\pm 2\mathbf{Q}\right) }\right\vert
_{T=T_{cP}}  \label{084}
\end{align}%
with $T_{n}\left( \mathbf{q},\mathbf{p},\mathbf{k}\right) =\left\langle
L_{n}^{-1}\left( \mathbf{q}\right) L_{n}^{-1}\left( \mathbf{p}\right)
L_{n}^{-1}\left( \mathbf{k}\right) \right\rangle /2$. The solution of the
system (\ref{080})-(\ref{081}) is found from%
\begin{equation}
\left\vert
\begin{array}{ccc}
P+t^{2}b_{-}+\delta _{-} & -t^{2}c_{-} & 0 \\
-t^{2}c_{-} & P+t^{2}a & -t^{2}c_{+} \\
0 & -t^{2}c_{+} & P+t^{2}b_{+}+\delta _{+}%
\end{array}%
\right\vert =0.
\end{equation}%
For $T>T^{\ast }$, when $q=0$, $\Delta _{\mathbf{+}2}=\Delta _{-2}$, which
makes it possible to write the solution in the form
\begin{equation}
T_{c}=T_{cP}\left[ 1-AS^{\pm }\left( \mathbf{Q}\right) \right]  \label{086}
\end{equation}%
with%
\begin{multline}
S^{\pm }\left( \mathbf{Q}\right) \equiv \frac{\left( a+b_{\pm }\right)
t^{2}+\delta _{\pm }}{2} \\
+\frac{t^{2}}{2}\sqrt{\left[ a-b_{\pm }-\delta _{\pm }/t^{2}\right]
^{2}+4c_{\pm }\sum\limits_{\pm }c_{\pm }}.  \label{087}
\end{multline}%
If $\sqrt{tT_{c0}}\ll \hbar v_{F}Q$ then it further simplifies, $S^{\pm
}\left( \mathbf{Q}\right) =at^{2}$. In Eq. (\ref{086}) those values of $\pm $
are chosen that maximize the critical temperature. In general case, if $%
H<H^{\ast }$ then within a second-order approximation in the small parameter
$t/T_{c0}$, $\Delta _{\mathbf{\pm }2}$ reads as
\begin{equation}
\Delta _{\mathbf{\pm }2}\approx \frac{t^{2}}{\left( \hbar v_{F}Q\right) ^{2}}%
\Delta _{0}  \label{088}
\end{equation}%
and the solution (\ref{086}) system (\ref{080}-\ref{081}) simplifies to (for
details see Appendix \ref{sec:appendix_C})%
\begin{equation}
P=\pi T_{cP}\sum\limits_{n}\frac{t^{2}}{\Omega _{n}^{3}}\left[ -1+\frac{1}{8}%
\frac{\left( \hbar v_{F}Q\right) ^{2}}{\Omega _{n}^{2}}+\frac{t^{2}}{\left(
\hbar v_{F}Q\right) ^{2}}\right] .  \label{089}
\end{equation}%
In the absence of the Zeeman effect
\begin{equation}
\ln \frac{T_{c}}{T_{ct}}=\frac{t^{2}}{\pi ^{2}T_{c0}^{2}}\left[ \frac{%
31\zeta \left( 5\right) }{128}\frac{\left( \hbar v_{F}Q\right) ^{2}}{\pi
^{2}T_{c0}^{2}}+\frac{7\zeta \left( 3\right) }{4}\frac{t^{2}}{\left( \hbar
v_{F}Q\right) ^{2}}\right] .  \label{090}
\end{equation}%
which is the same as Eq. (\ref{035}). Thus, within the expansion model (\ref%
{050}) we obtained the upper critical field versus the superconducting onset
temperature. This equation naturally describes the crossover between two
regimes: the Lowerence-Doniach phase and the beginning of the Lebed
re-entrant phase.

\begin{figure*}[tbp]
\resizebox{0.5\columnwidth}{!}{\rotatebox{0}{
\includegraphics{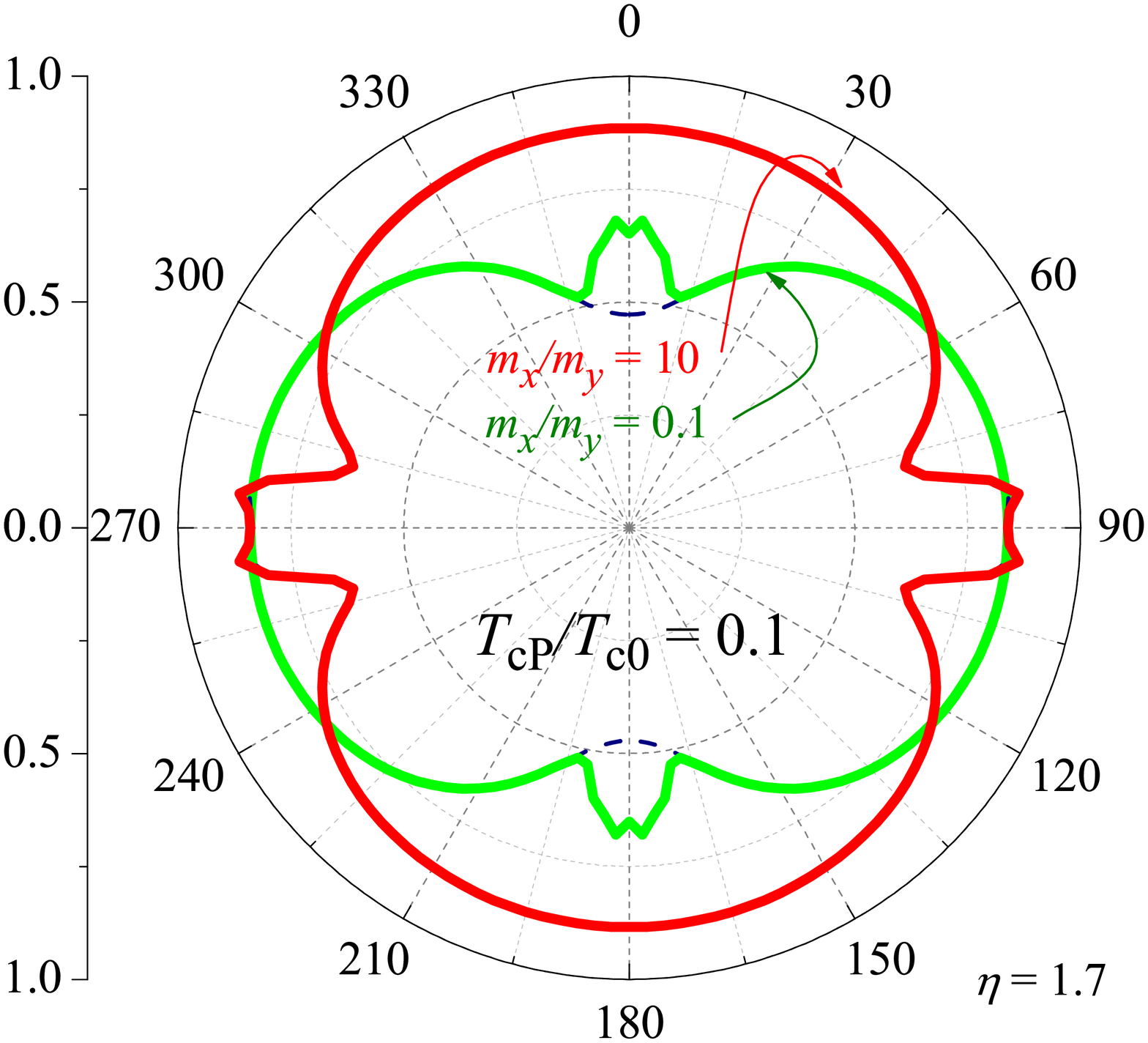}}}
\resizebox{0.5\columnwidth}{!}{\rotatebox{0}{
\includegraphics{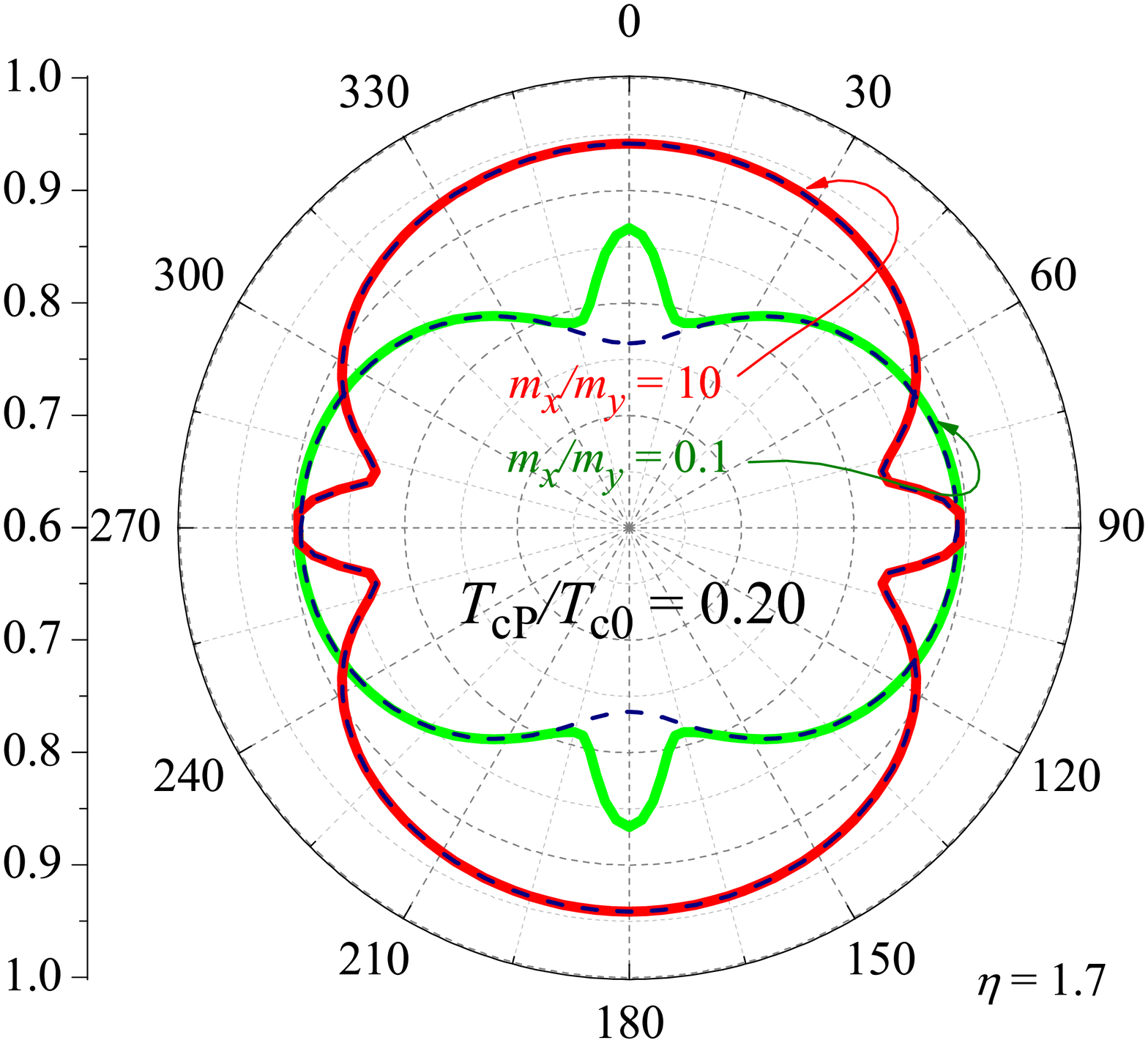}}}
\resizebox{0.5\columnwidth}{!}{\rotatebox{0}{
\includegraphics{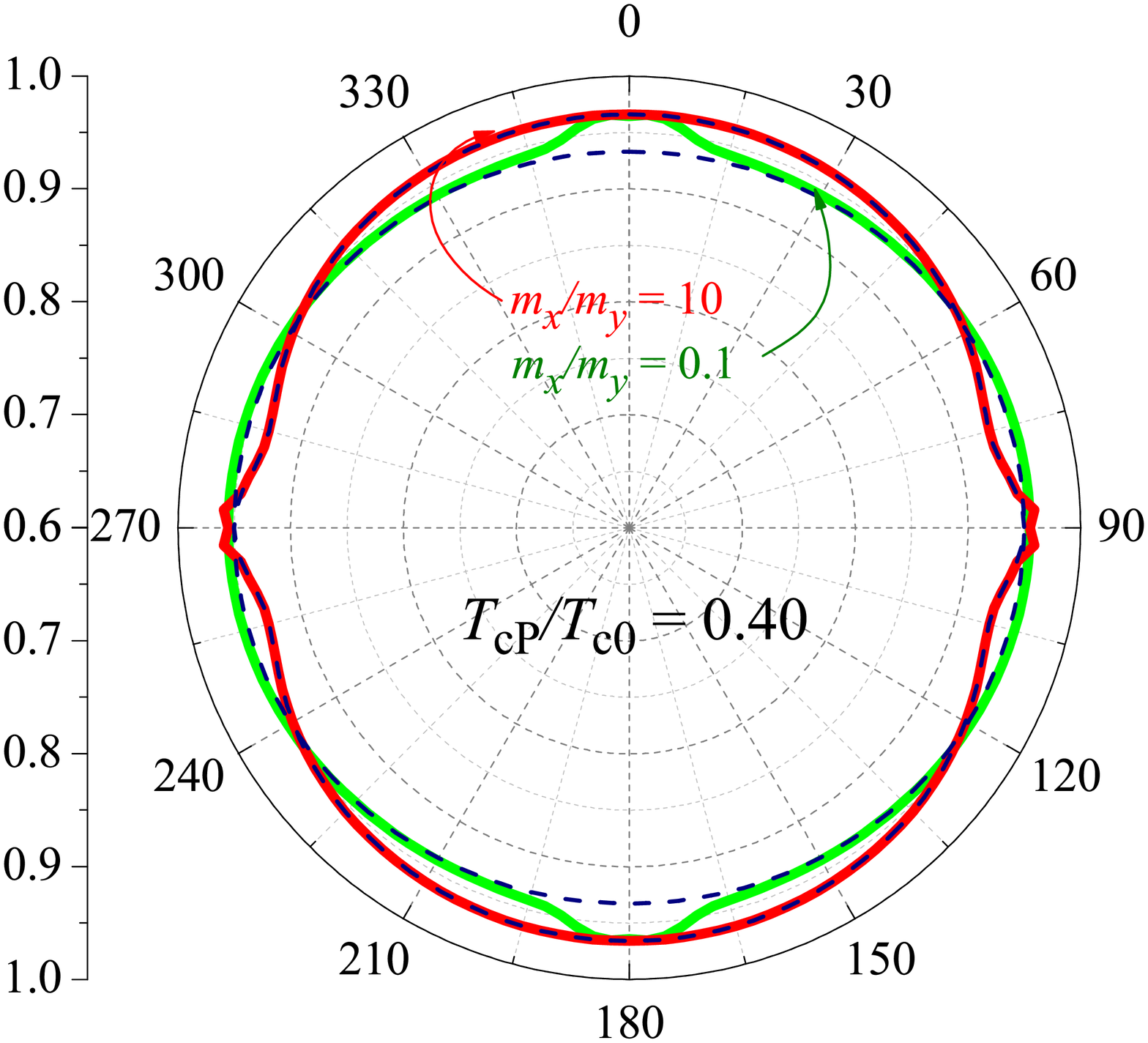}}}
\resizebox{0.5\columnwidth}{!}{\rotatebox{0}{
\includegraphics{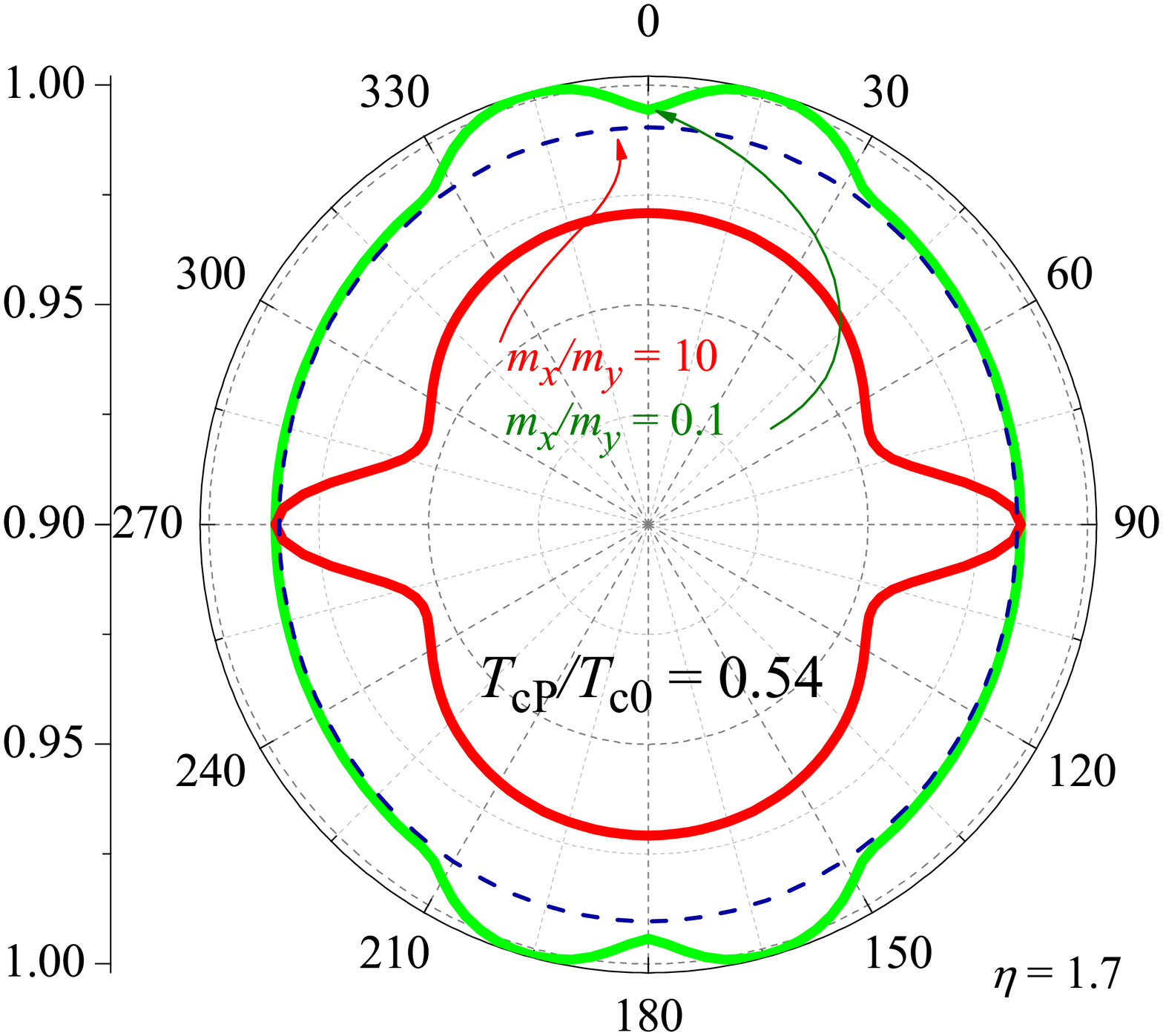}}}
\resizebox{0.5\columnwidth}{!}{\rotatebox{0}{
\includegraphics{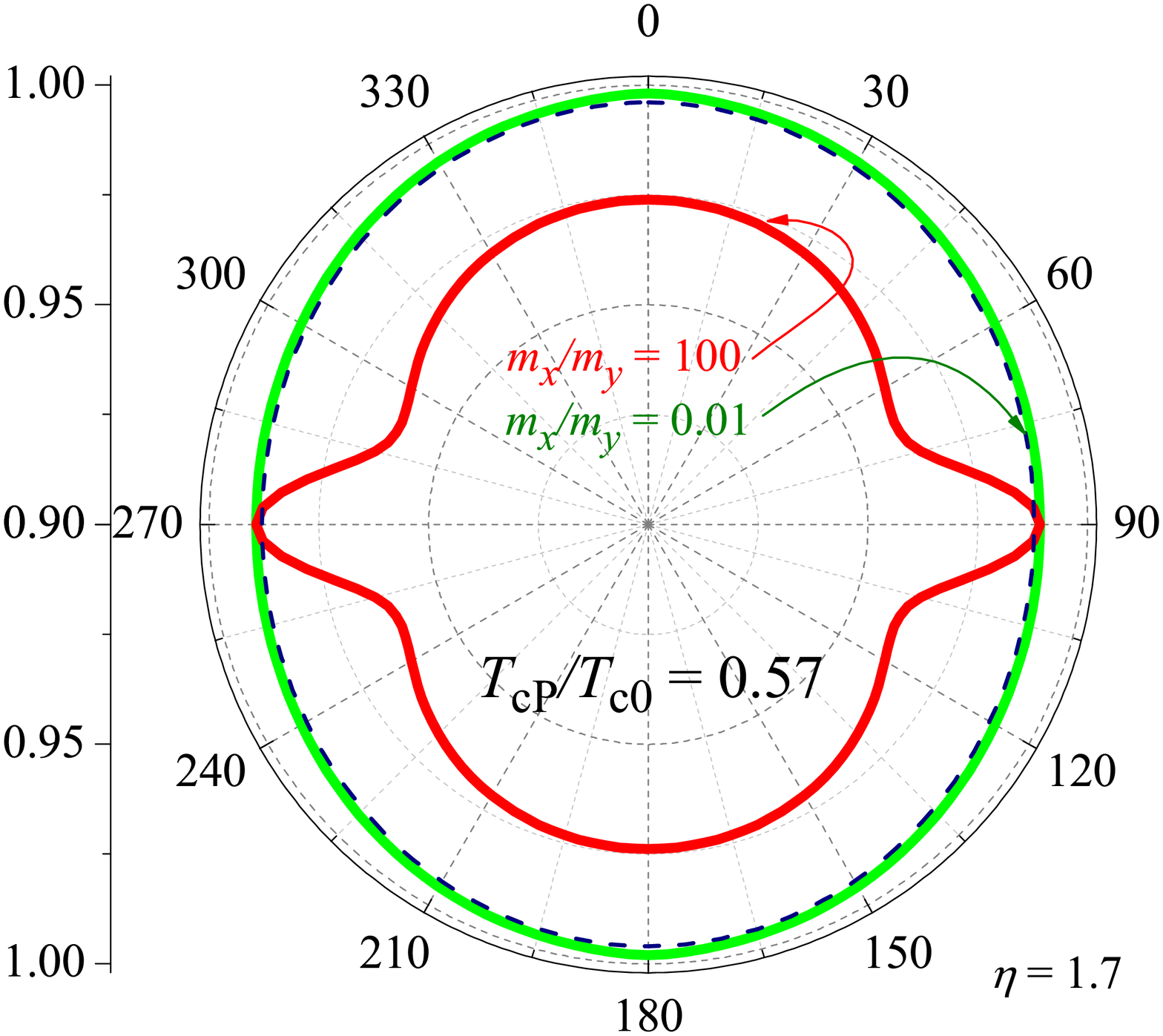}}}
\resizebox{0.5\columnwidth}{!}{\rotatebox{0}{
\includegraphics{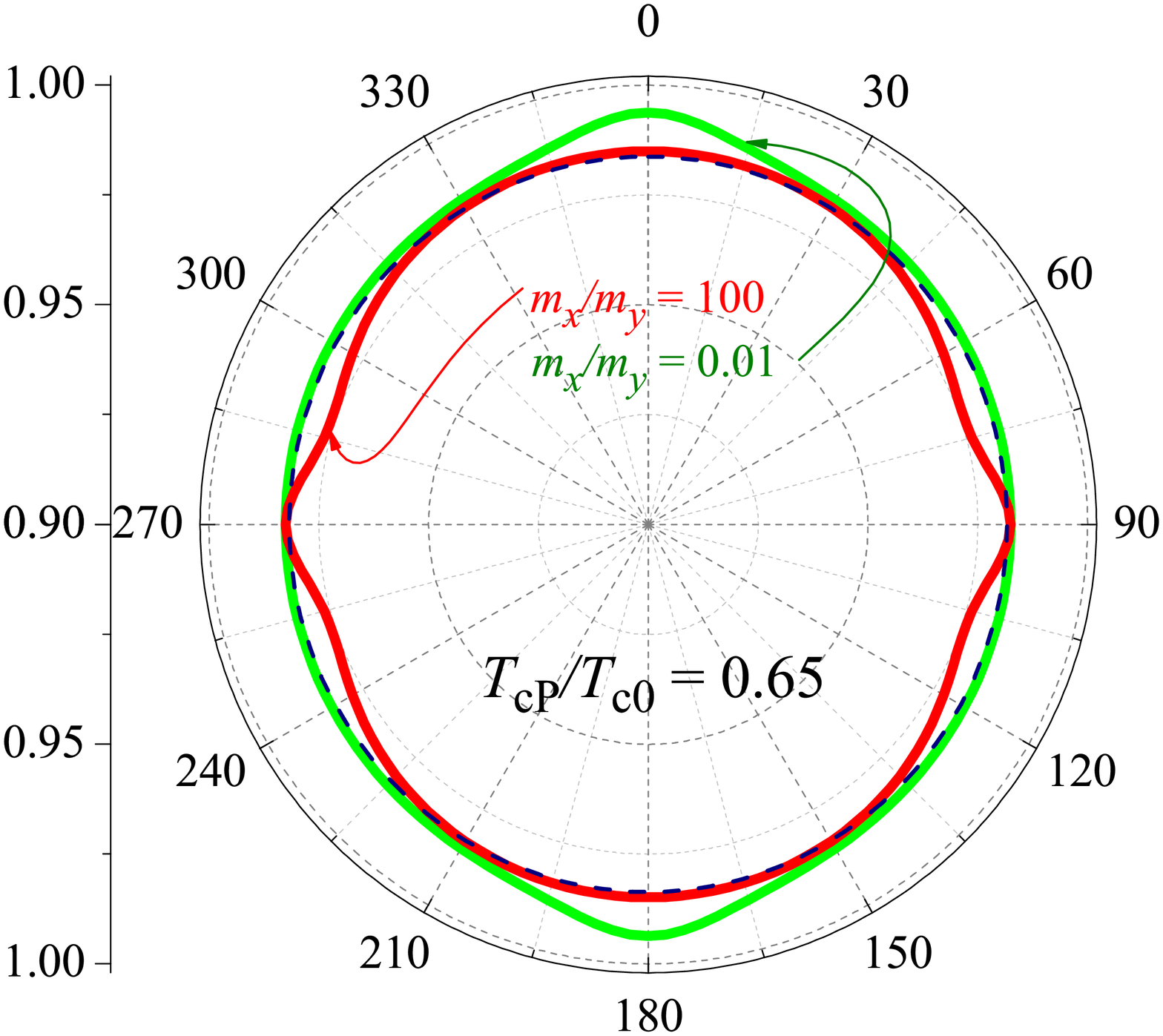}}}
\resizebox{0.5\columnwidth}{!}{\rotatebox{0}{
\includegraphics{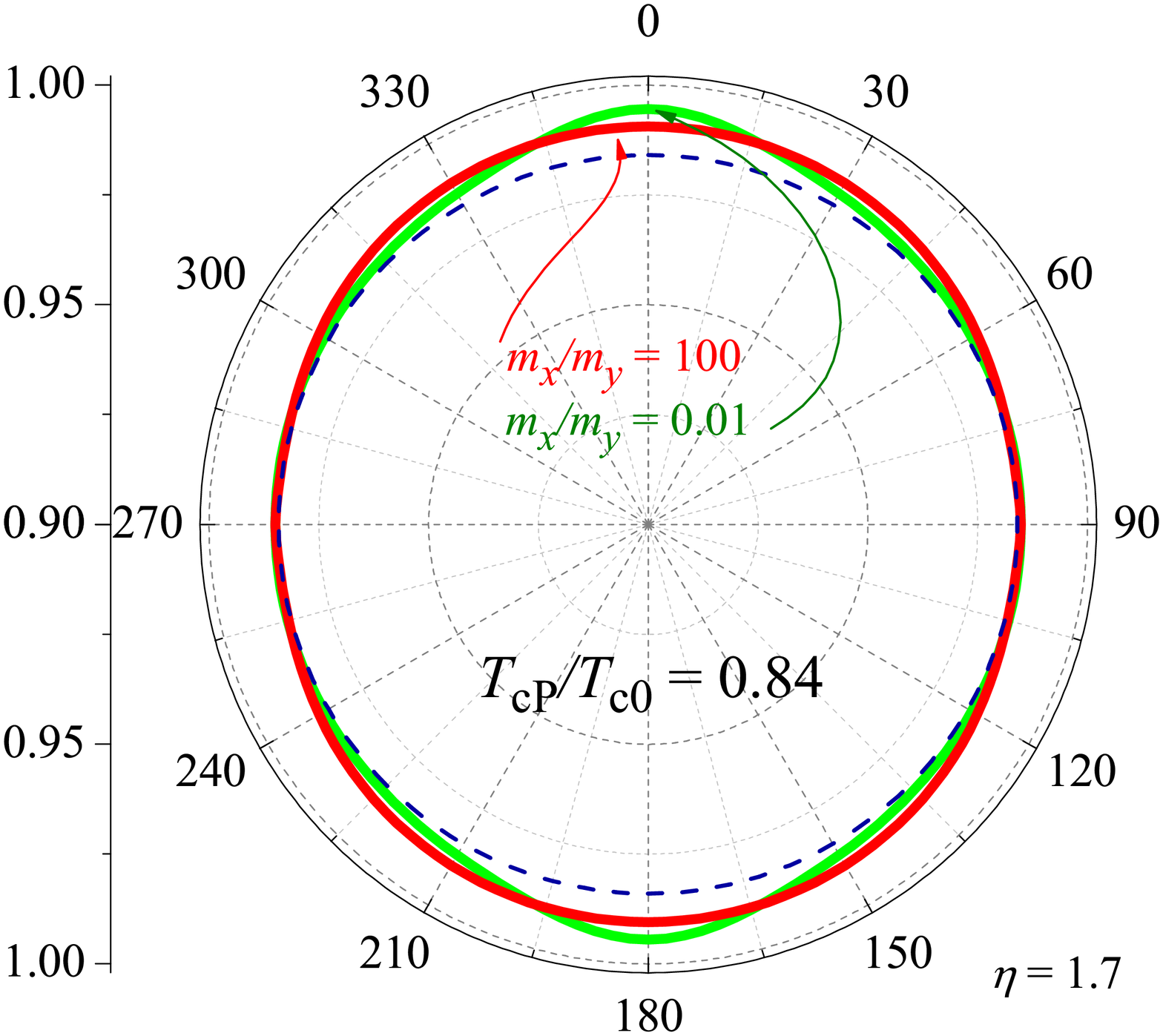}}}
\resizebox{0.5\columnwidth}{!}{\rotatebox{0}{
\includegraphics{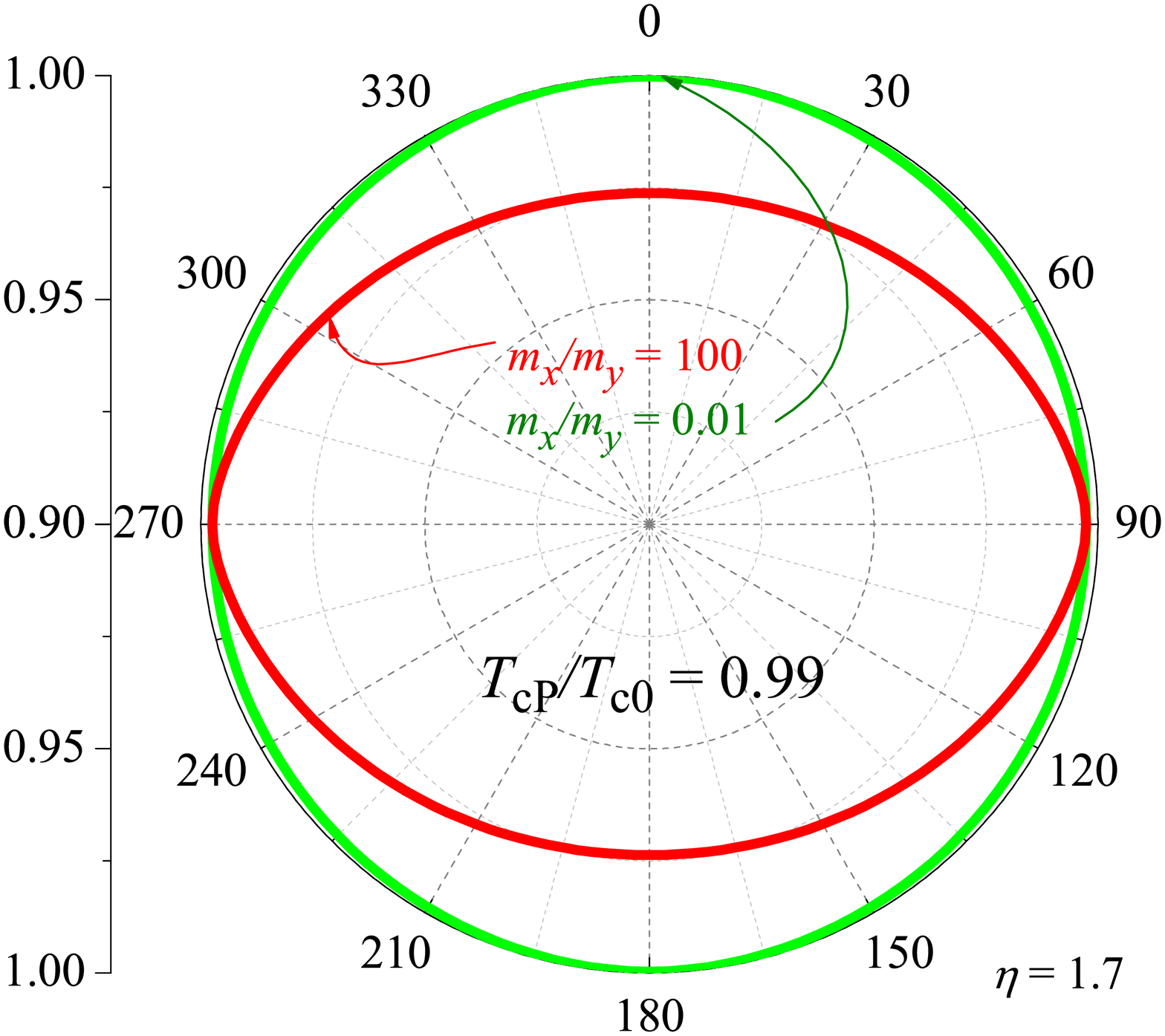}}}
\caption{Normalized transition temperature, $T_{c}\left( \protect\alpha %
\right) /T_{cP}$ as a function of $\protect\alpha $ for several $%
T_{cP}/T_{c0}$, $t/T_{c0}=0.25$, and $\protect\eta =1.7$. For the purpose of
clarity the shown range of $T_{c}\left( \protect\alpha \right) /T_{cP}$ is
from 0.9 till 1.0. Dashed lines are for $\Delta _{\mathbf{\pm }2}\neq 0$.}
\label{fig.06}
\end{figure*}
\begin{figure*}[tbp]
\resizebox{0.5\columnwidth}{!}{\rotatebox{0}{
\includegraphics{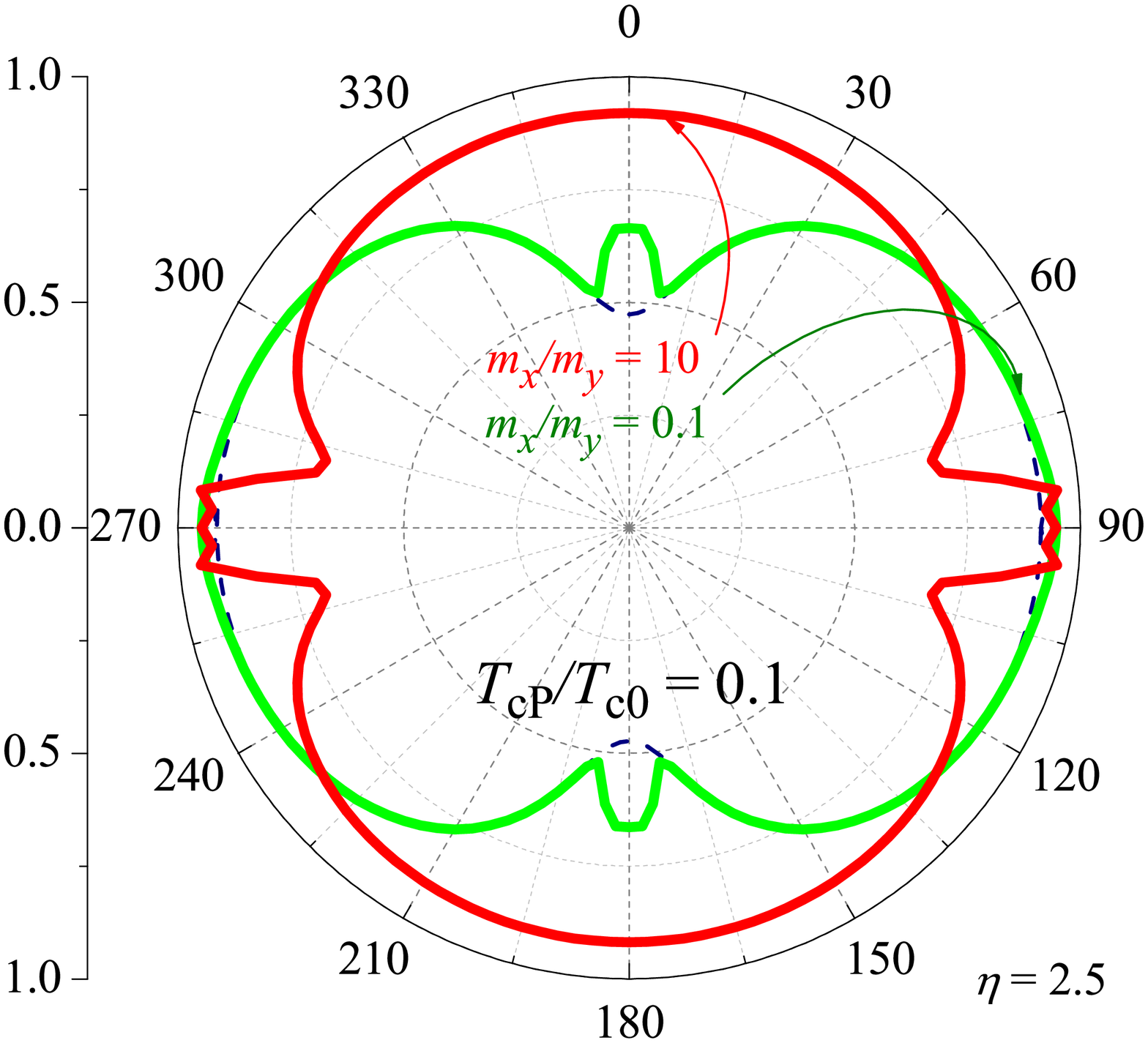}}}
\resizebox{0.5\columnwidth}{!}{\rotatebox{0}{
\includegraphics{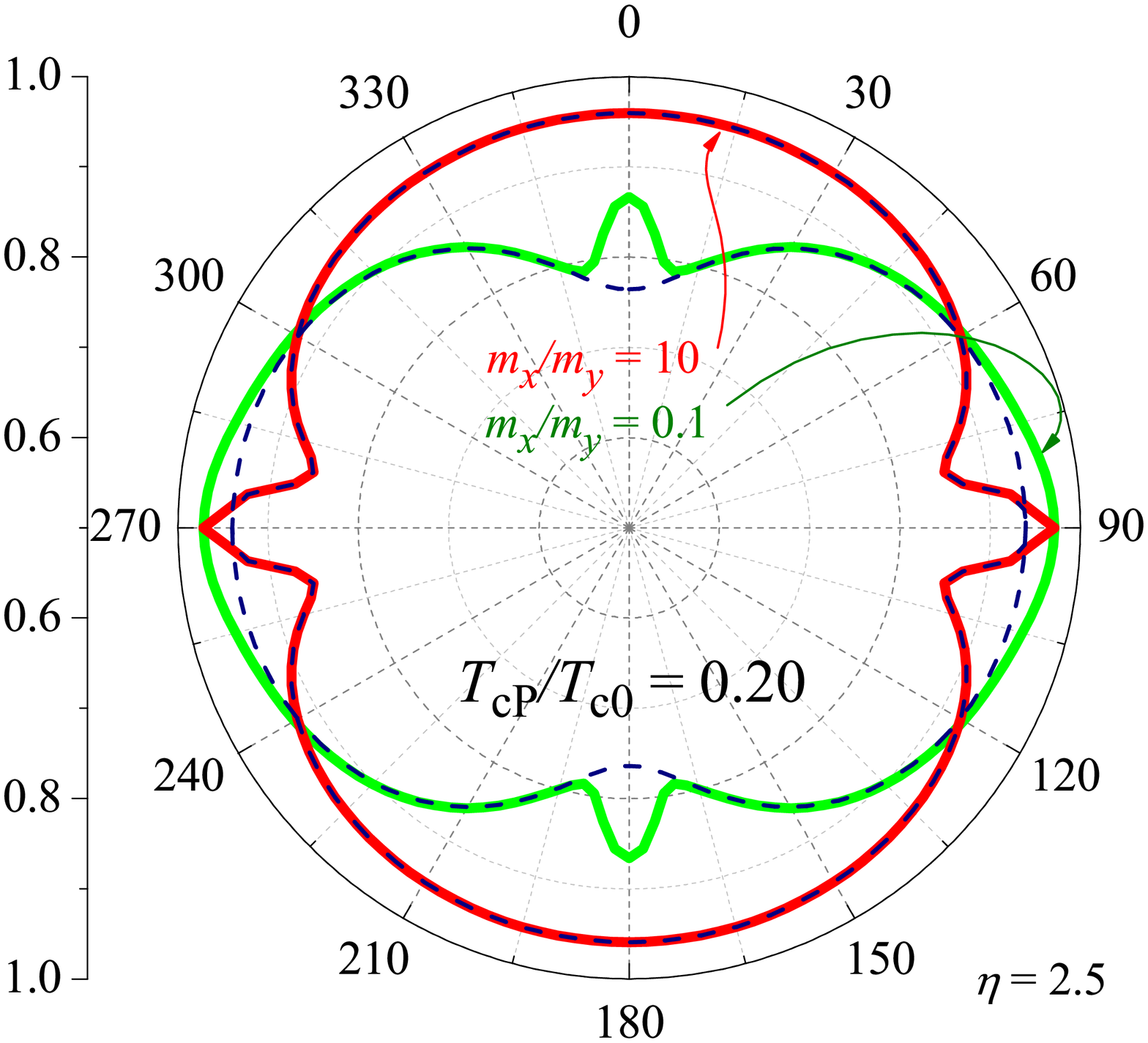}}}
\resizebox{0.5\columnwidth}{!}{\rotatebox{0}{
\includegraphics{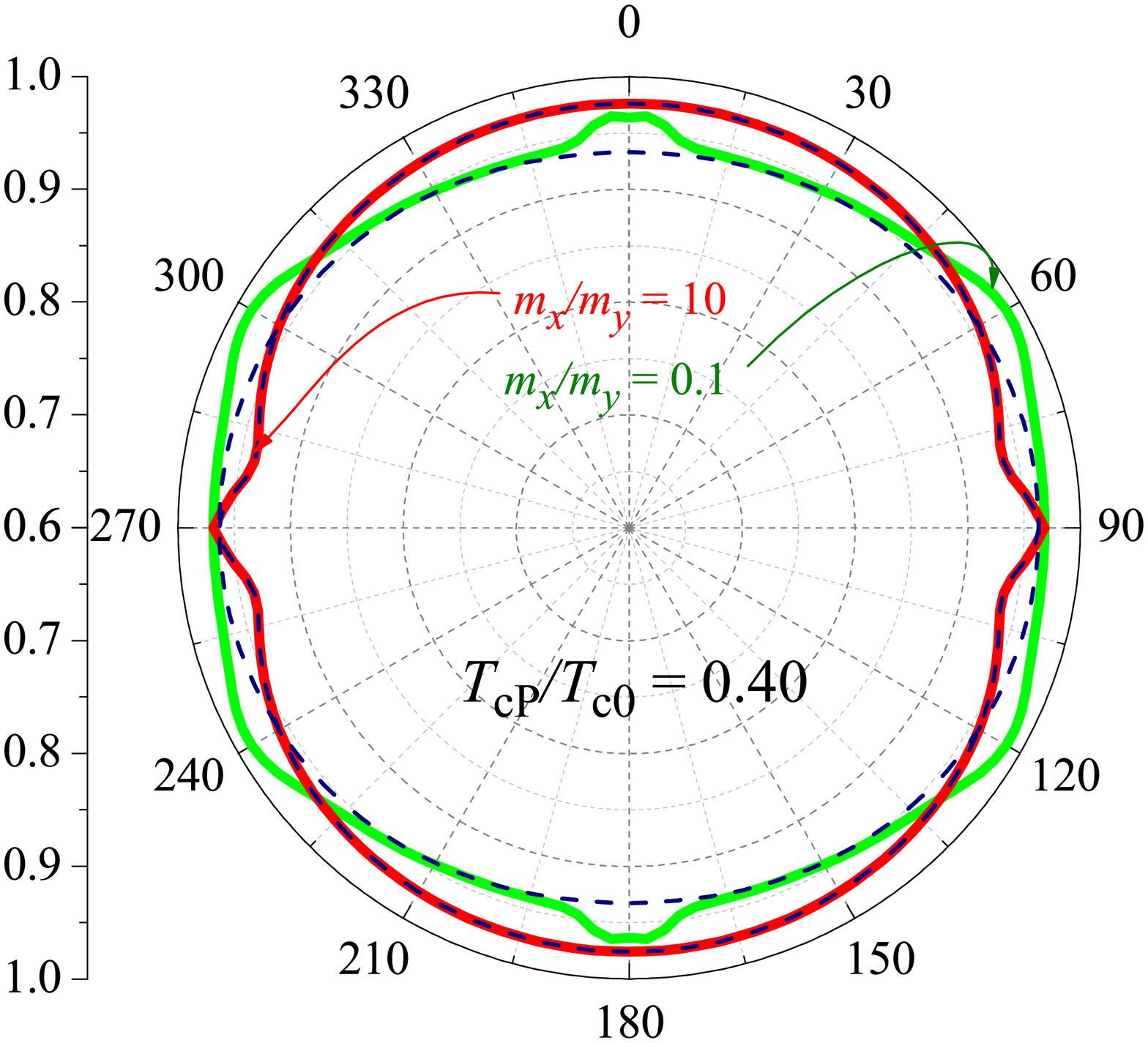}}}
\resizebox{0.5\columnwidth}{!}{\rotatebox{0}{
\includegraphics{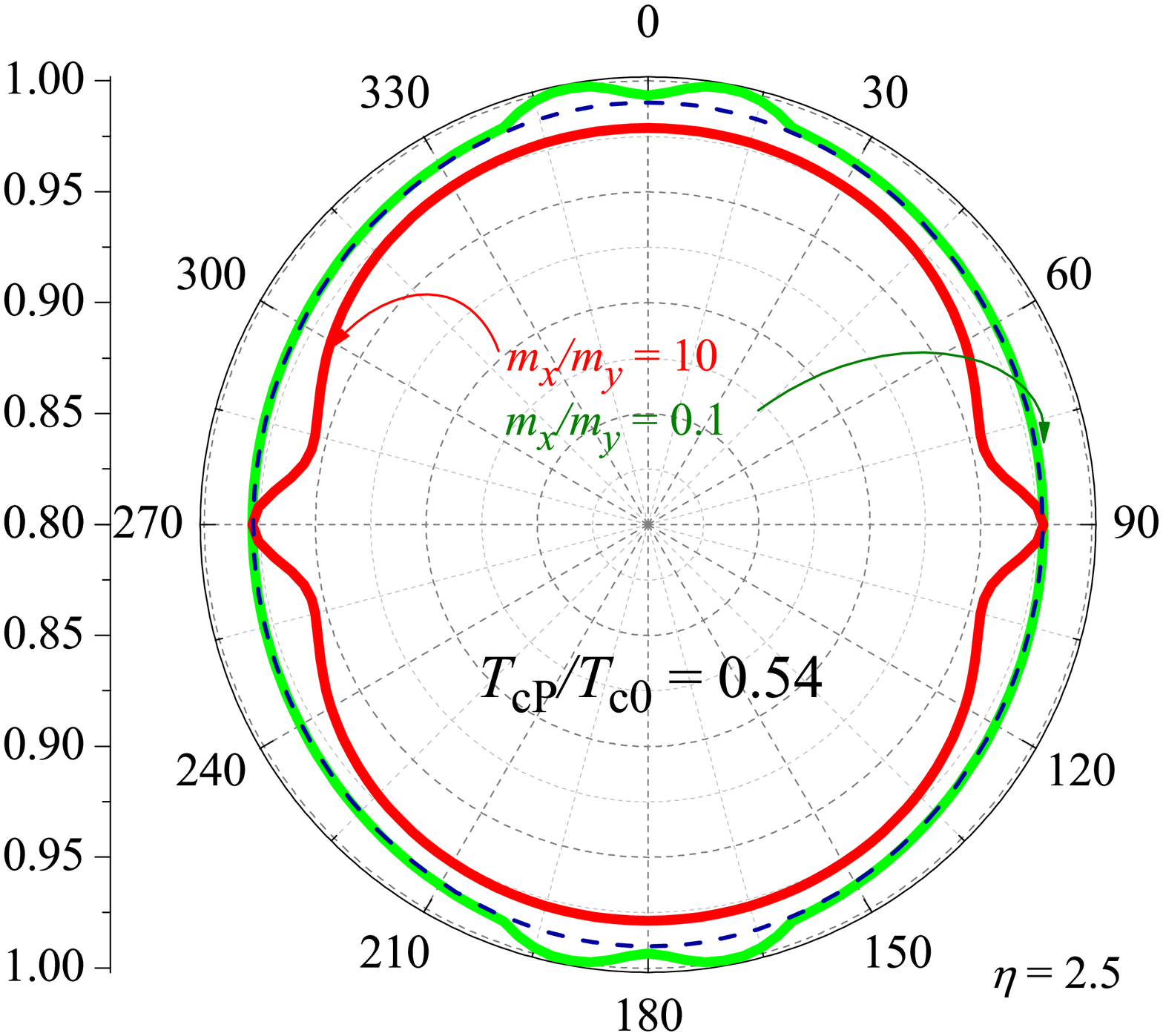}}}
\resizebox{0.5\columnwidth}{!}{\rotatebox{0}{
\includegraphics{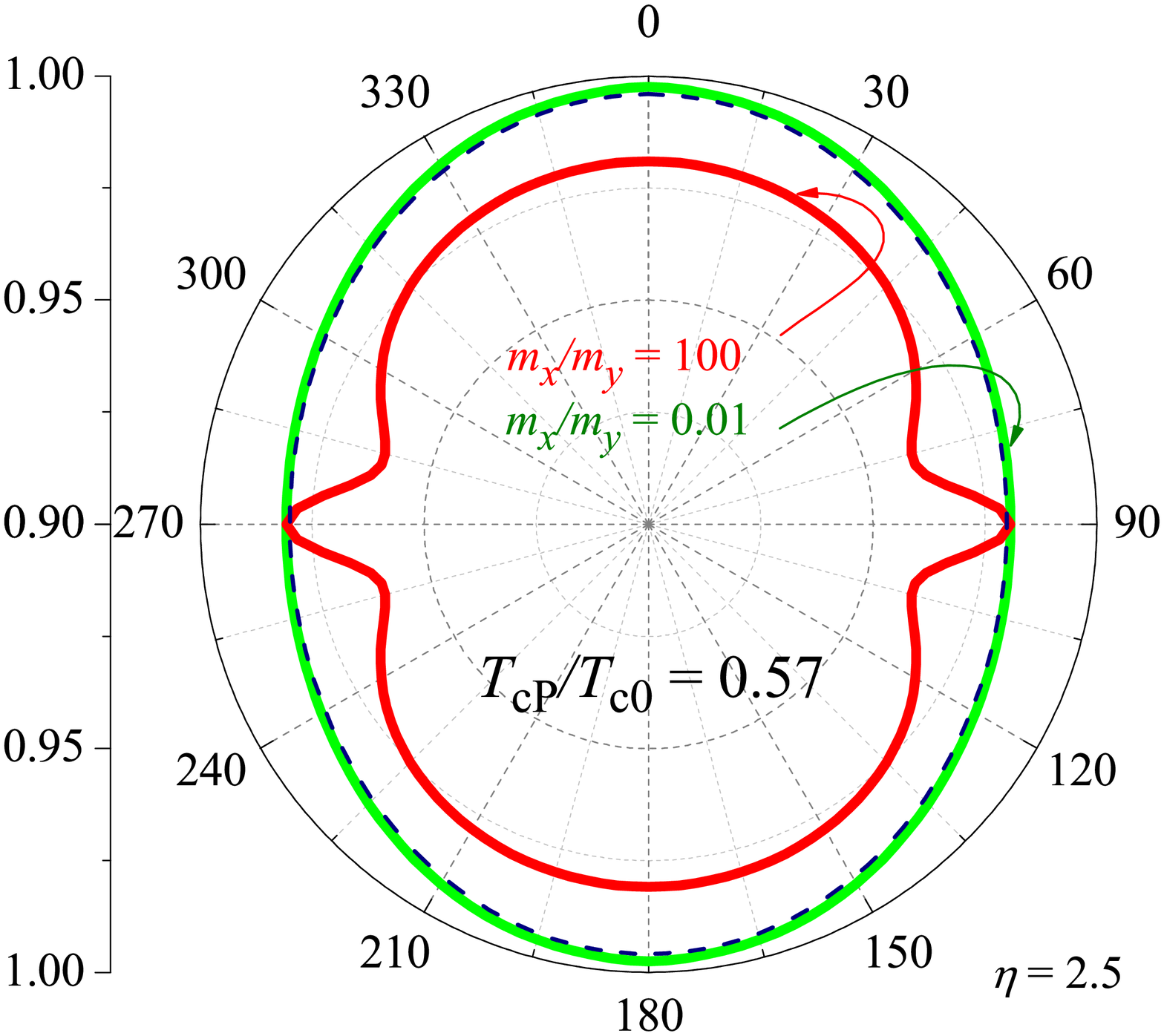}}}
\resizebox{0.5\columnwidth}{!}{\rotatebox{0}{
\includegraphics{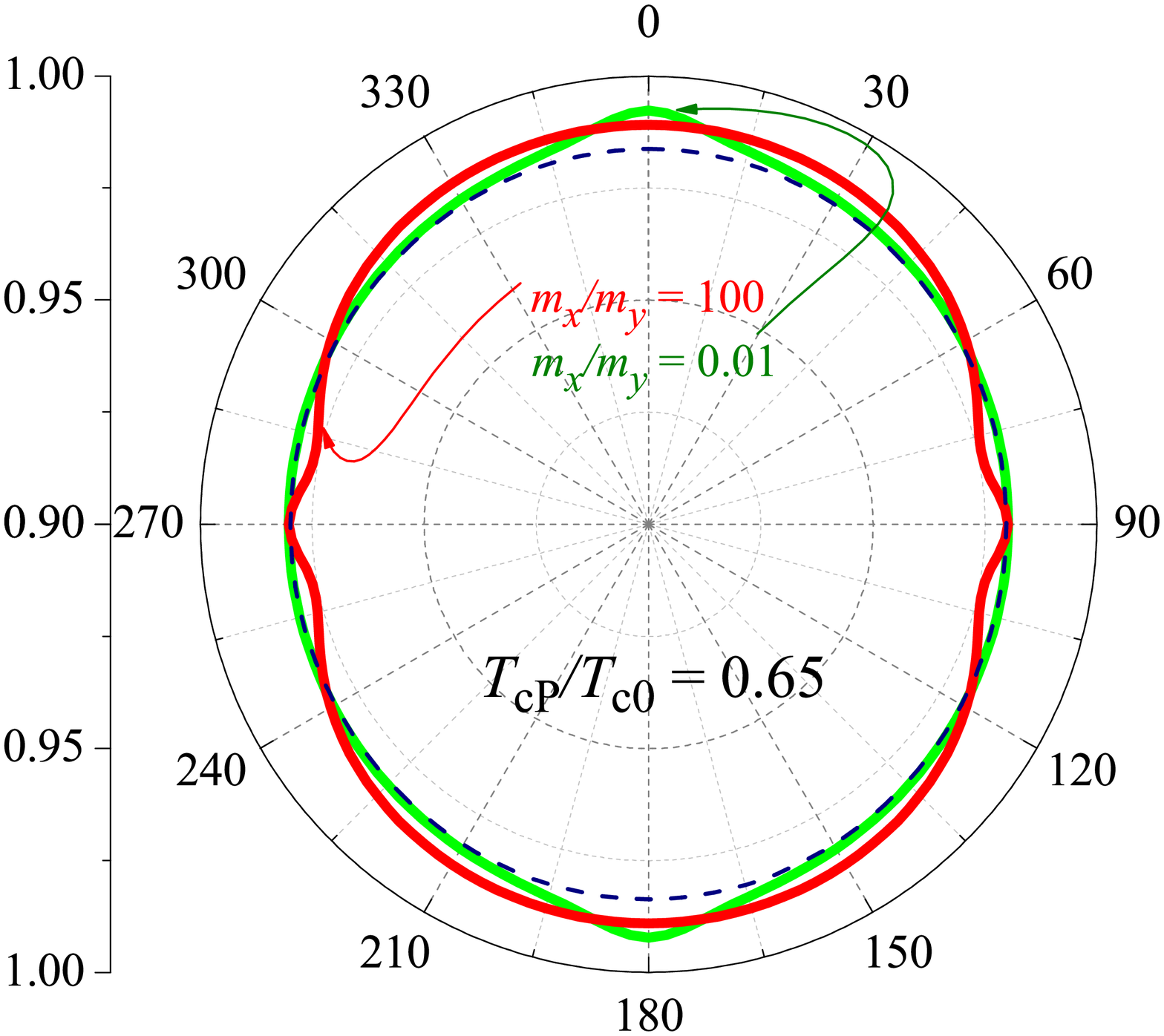}}}
\resizebox{0.5\columnwidth}{!}{\rotatebox{0}{
\includegraphics{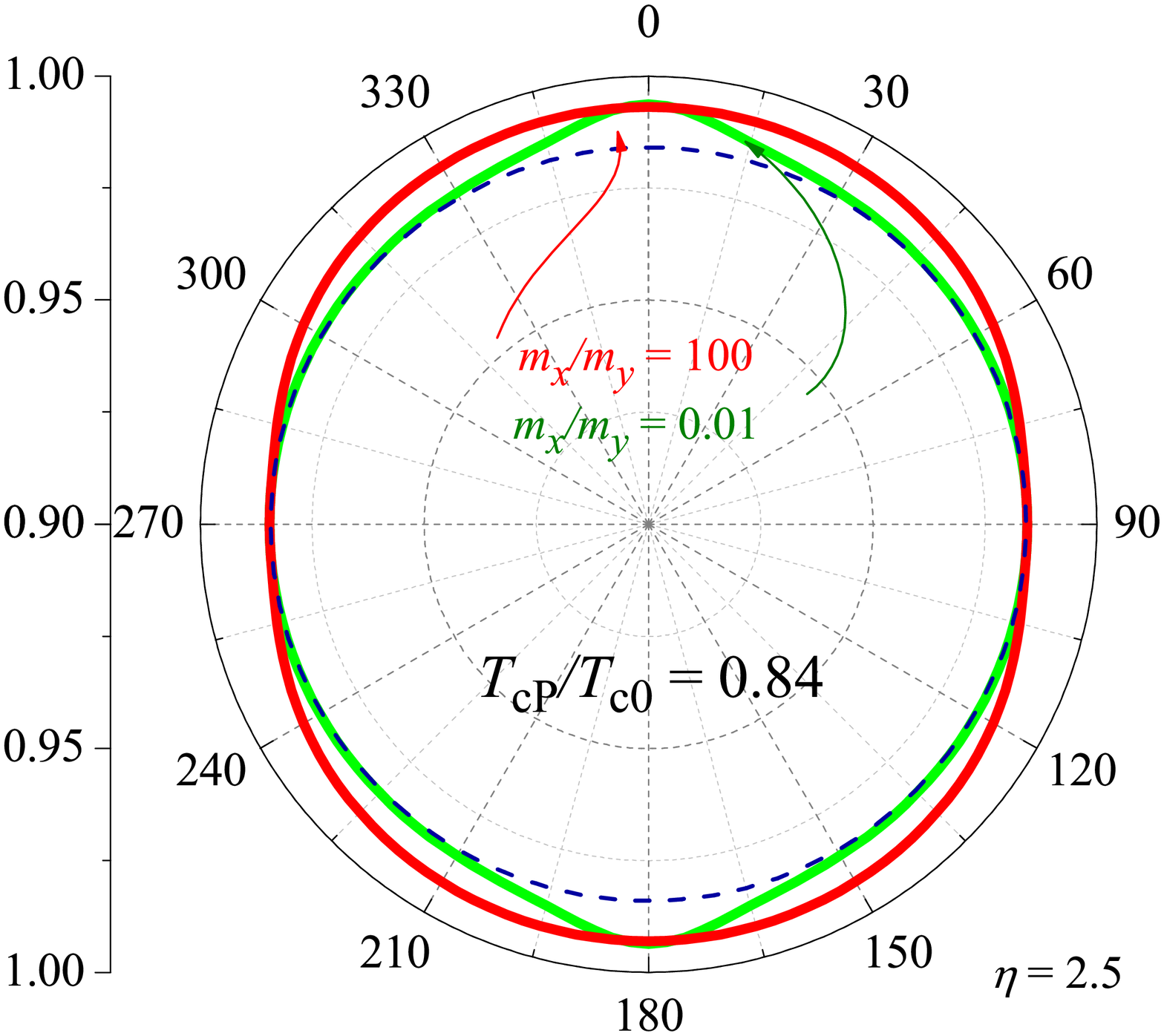}}}
\resizebox{0.5\columnwidth}{!}{\rotatebox{0}{
\includegraphics{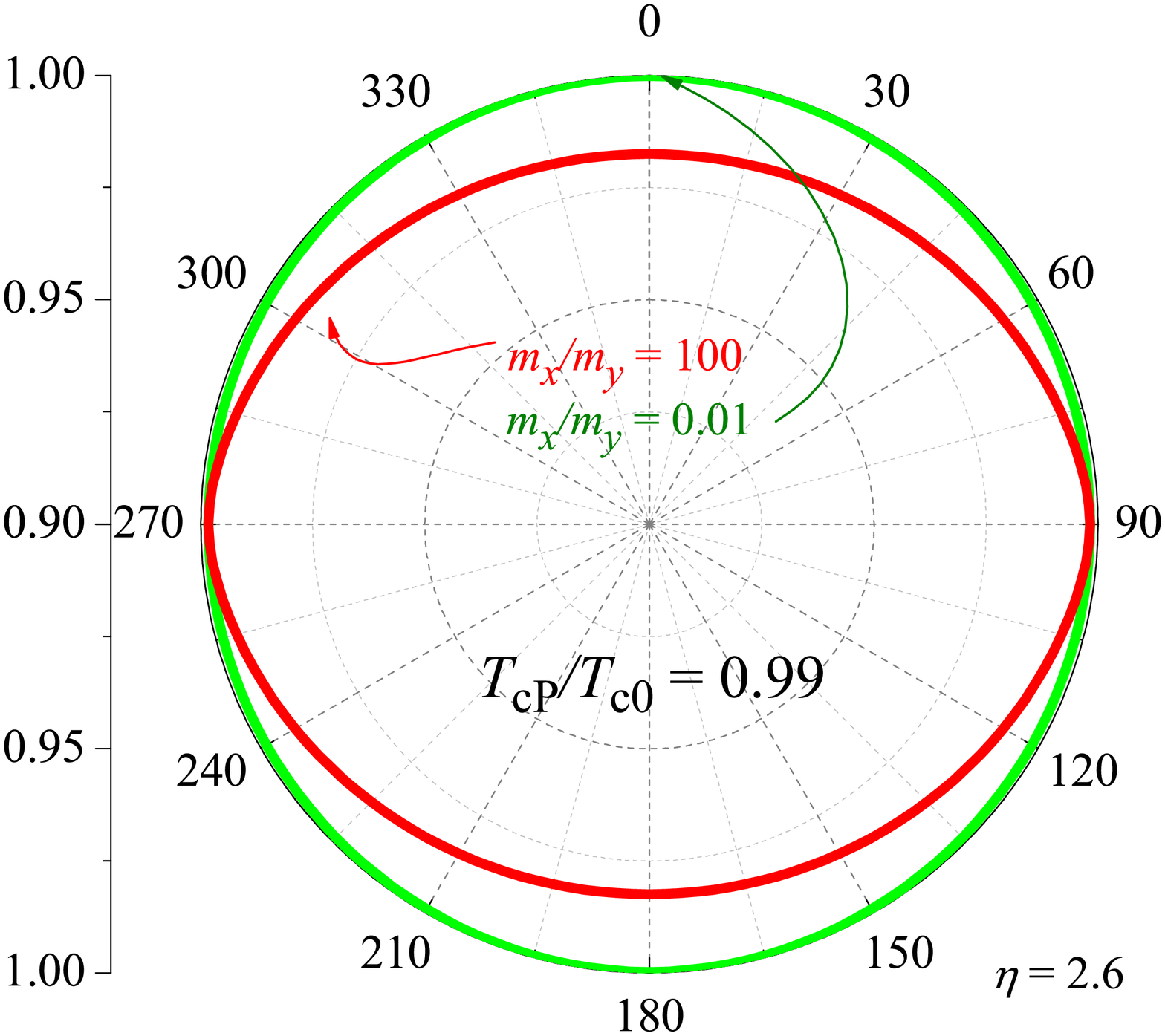}}}
\caption{The same as in Fig. \protect\ref{fig.05} but calculated for $%
\protect\eta =2.5$, which corresponds to $v_{F}=7.5\times 10^{4}~\mathrm{%
m/\sec }$. }
\label{fig.07}
\end{figure*}
\begin{figure*}[tbp]
\resizebox{0.5\columnwidth}{!}{\rotatebox{0}{
\includegraphics{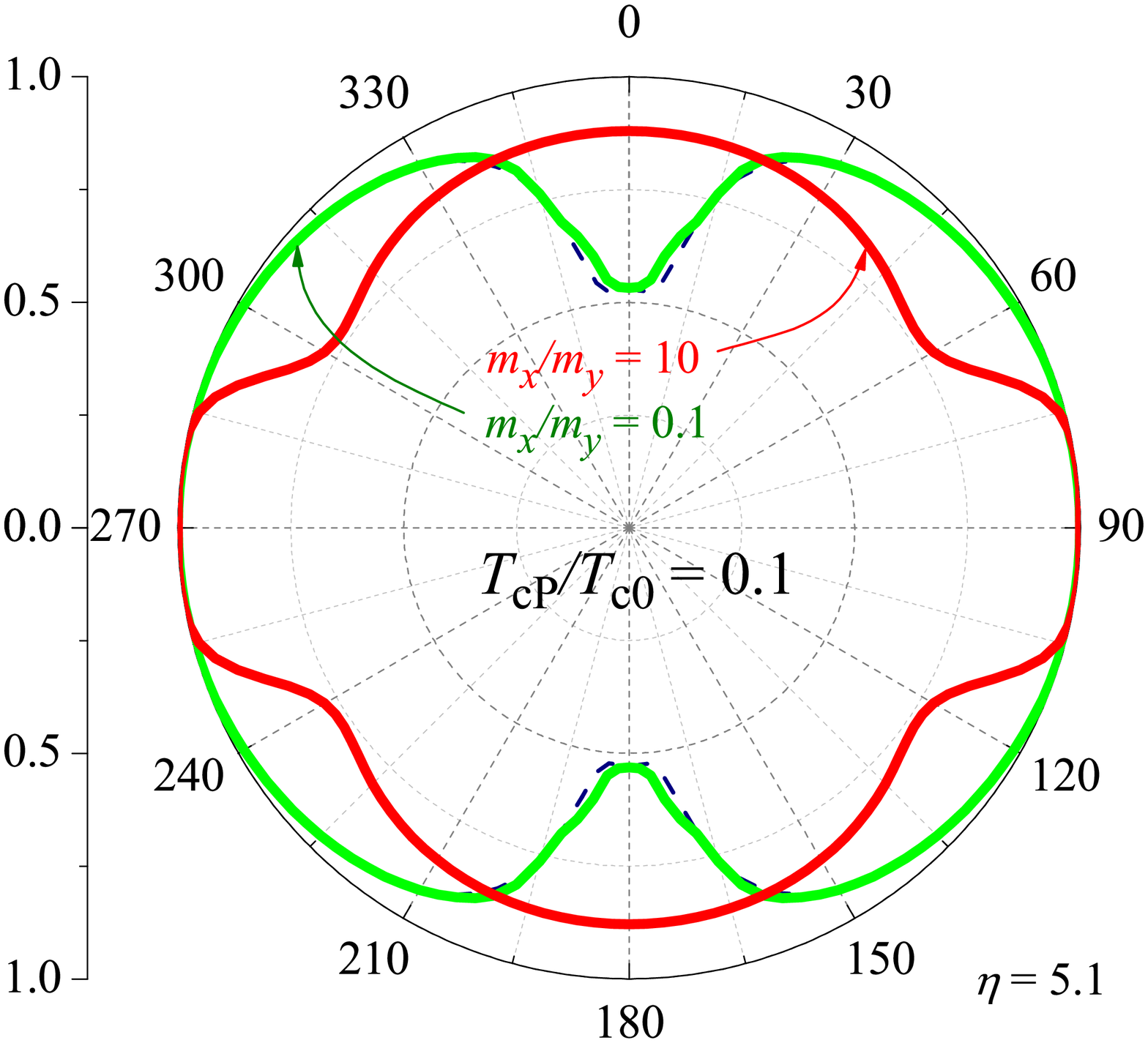}}}
\resizebox{0.5\columnwidth}{!}{\rotatebox{0}{
\includegraphics{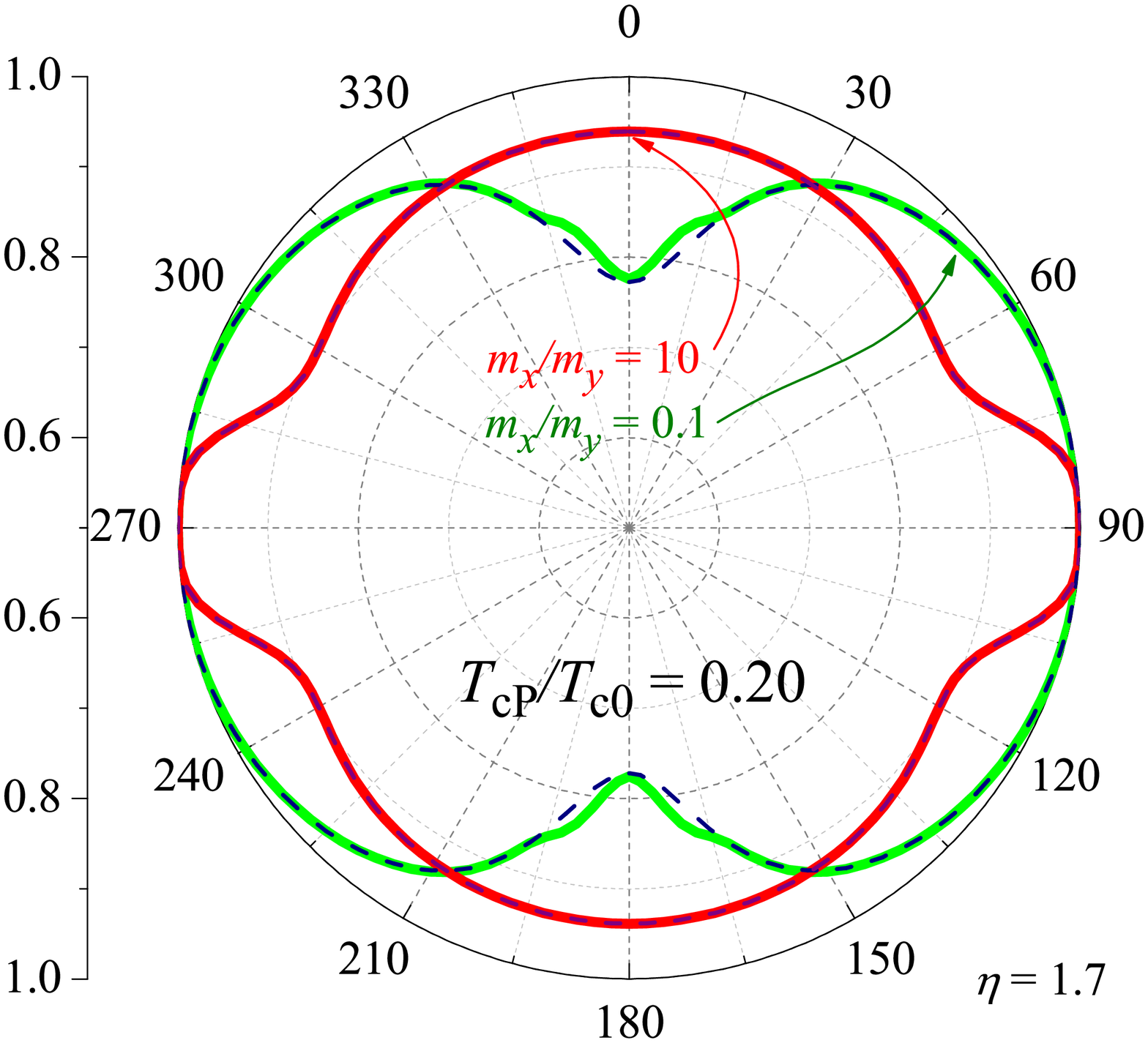}}}
\resizebox{0.5\columnwidth}{!}{\rotatebox{0}{
\includegraphics{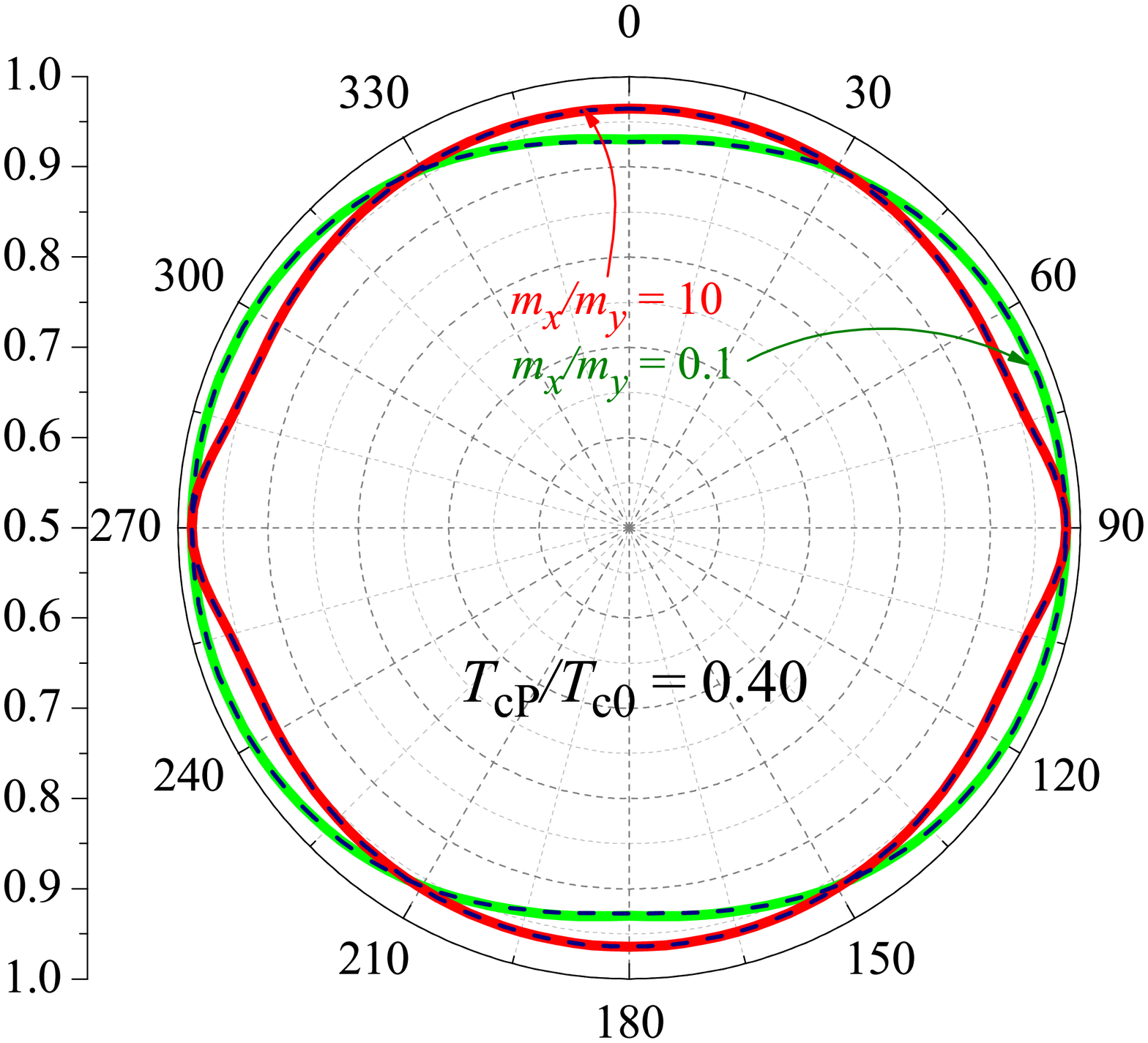}}}
\resizebox{0.5\columnwidth}{!}{\rotatebox{0}{
\includegraphics{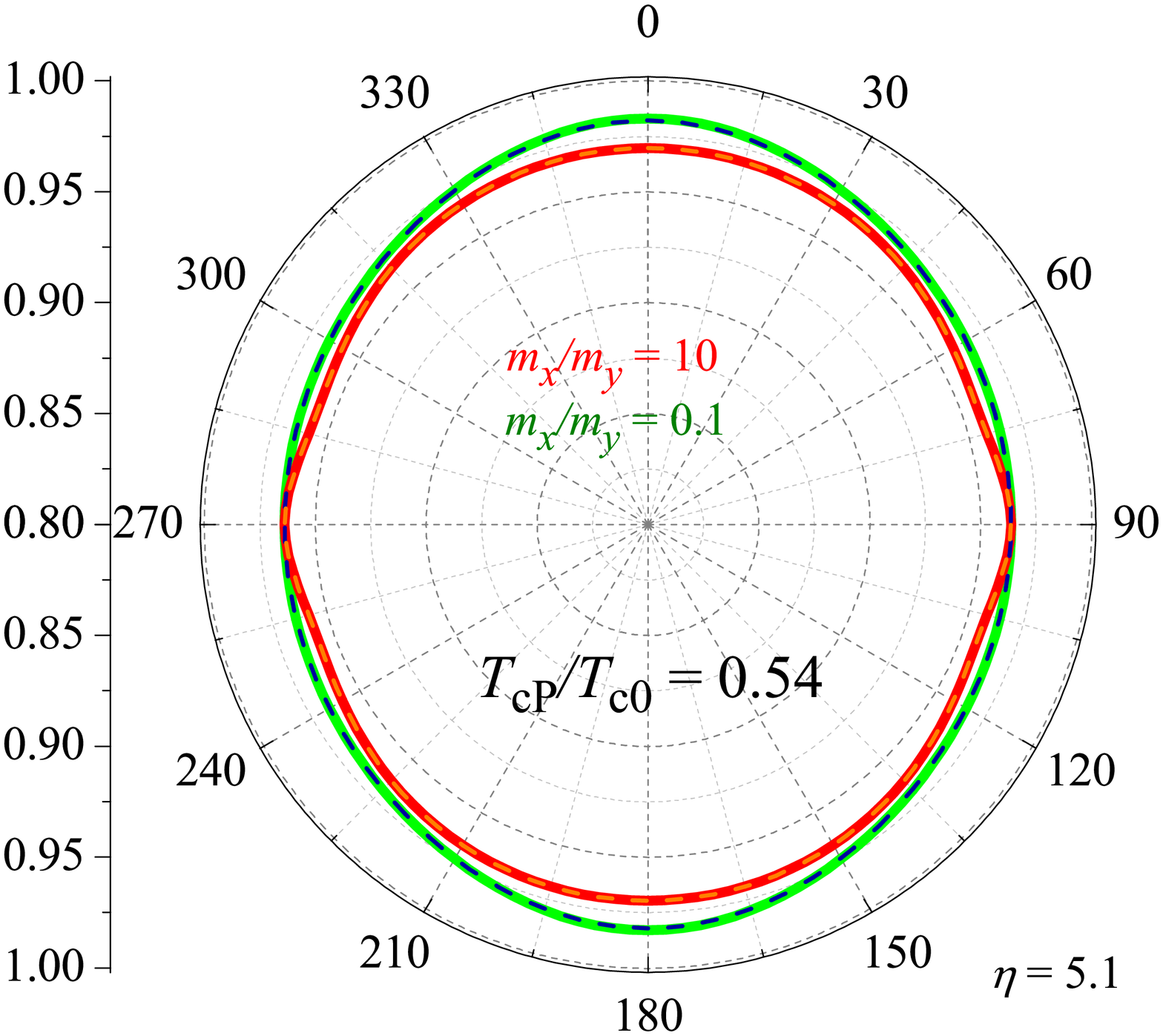}}}
\resizebox{0.5\columnwidth}{!}{\rotatebox{0}{
\includegraphics{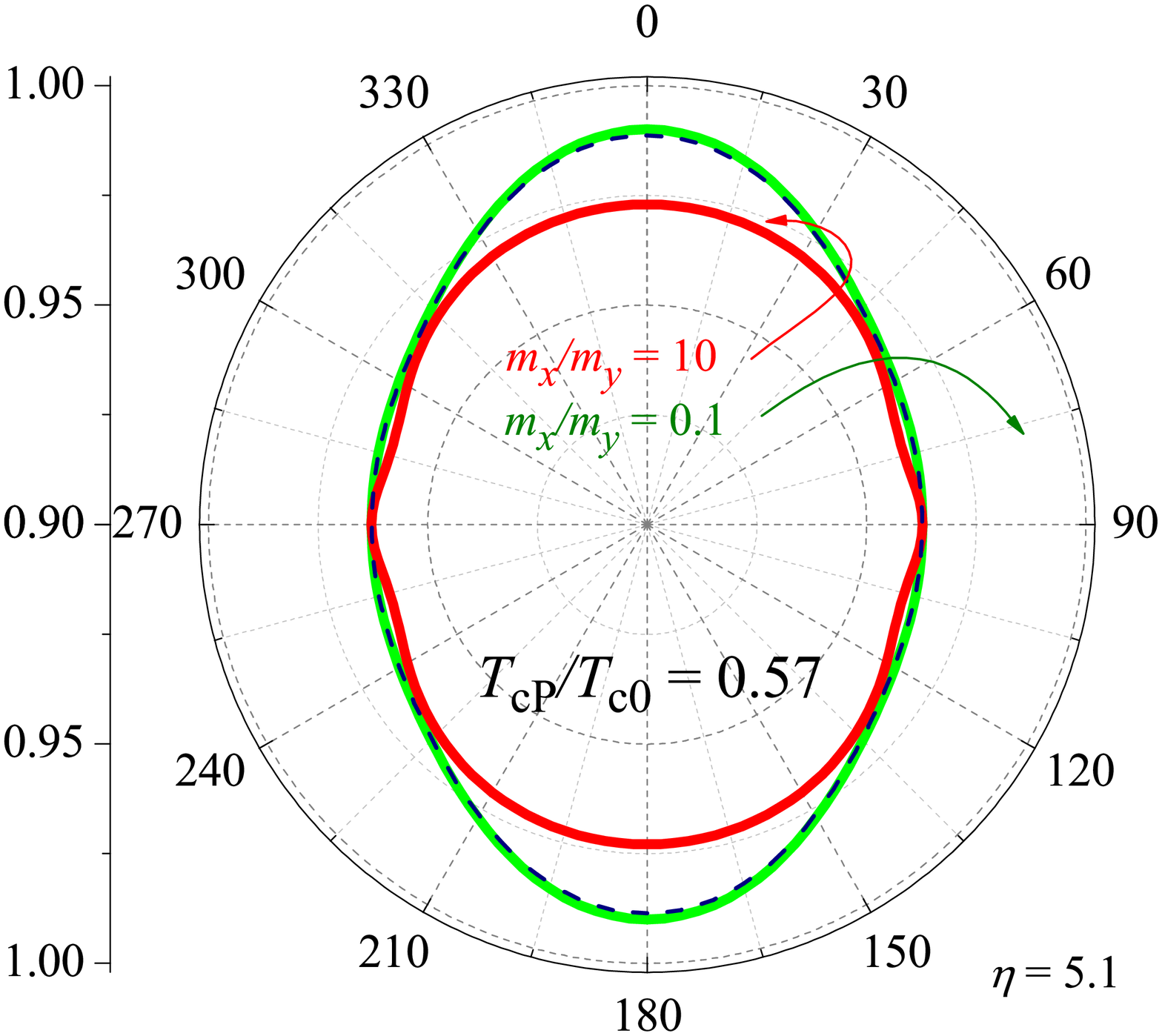}}}
\resizebox{0.5\columnwidth}{!}{\rotatebox{0}{
\includegraphics{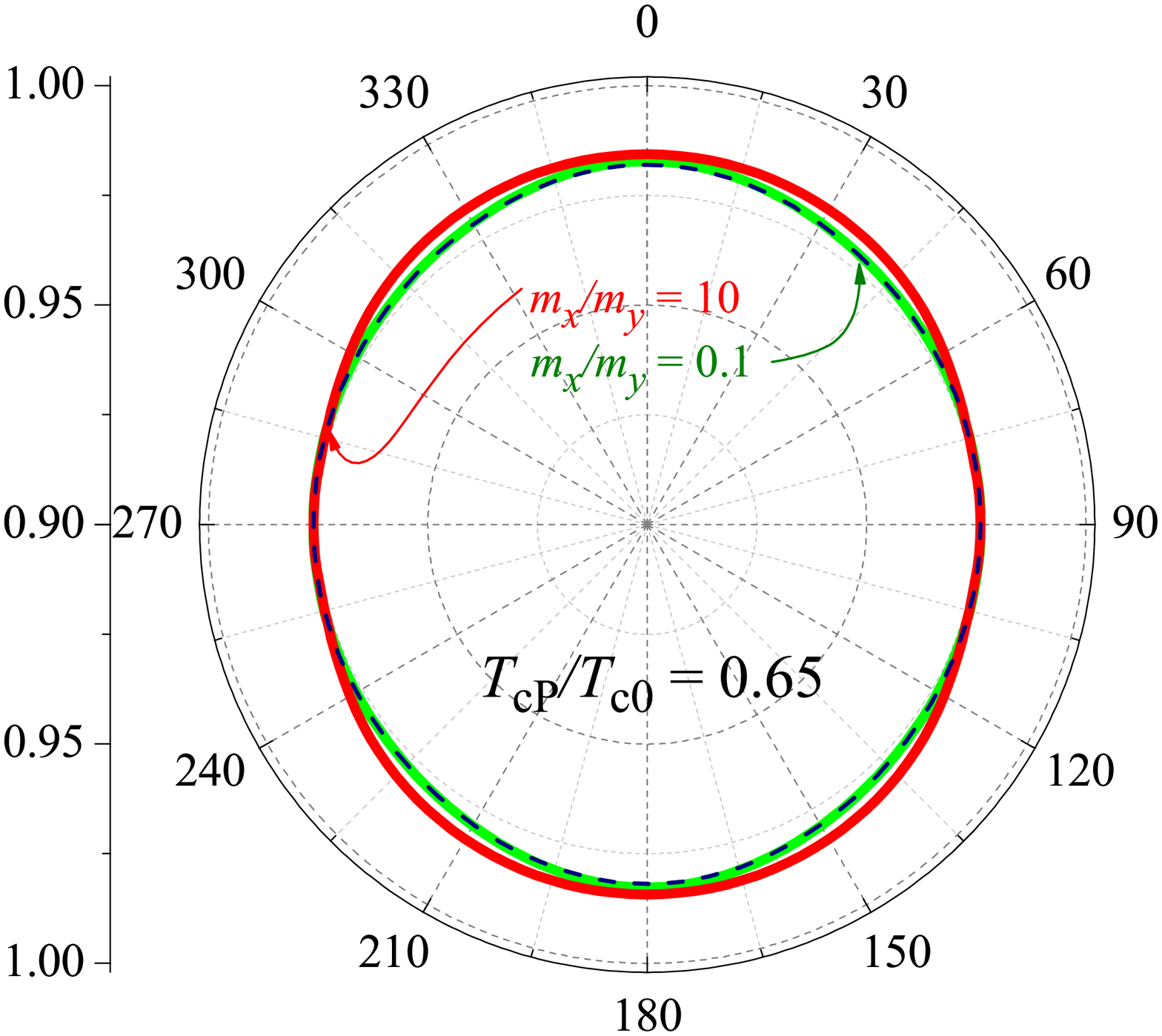}}}
\resizebox{0.5\columnwidth}{!}{\rotatebox{0}{
\includegraphics{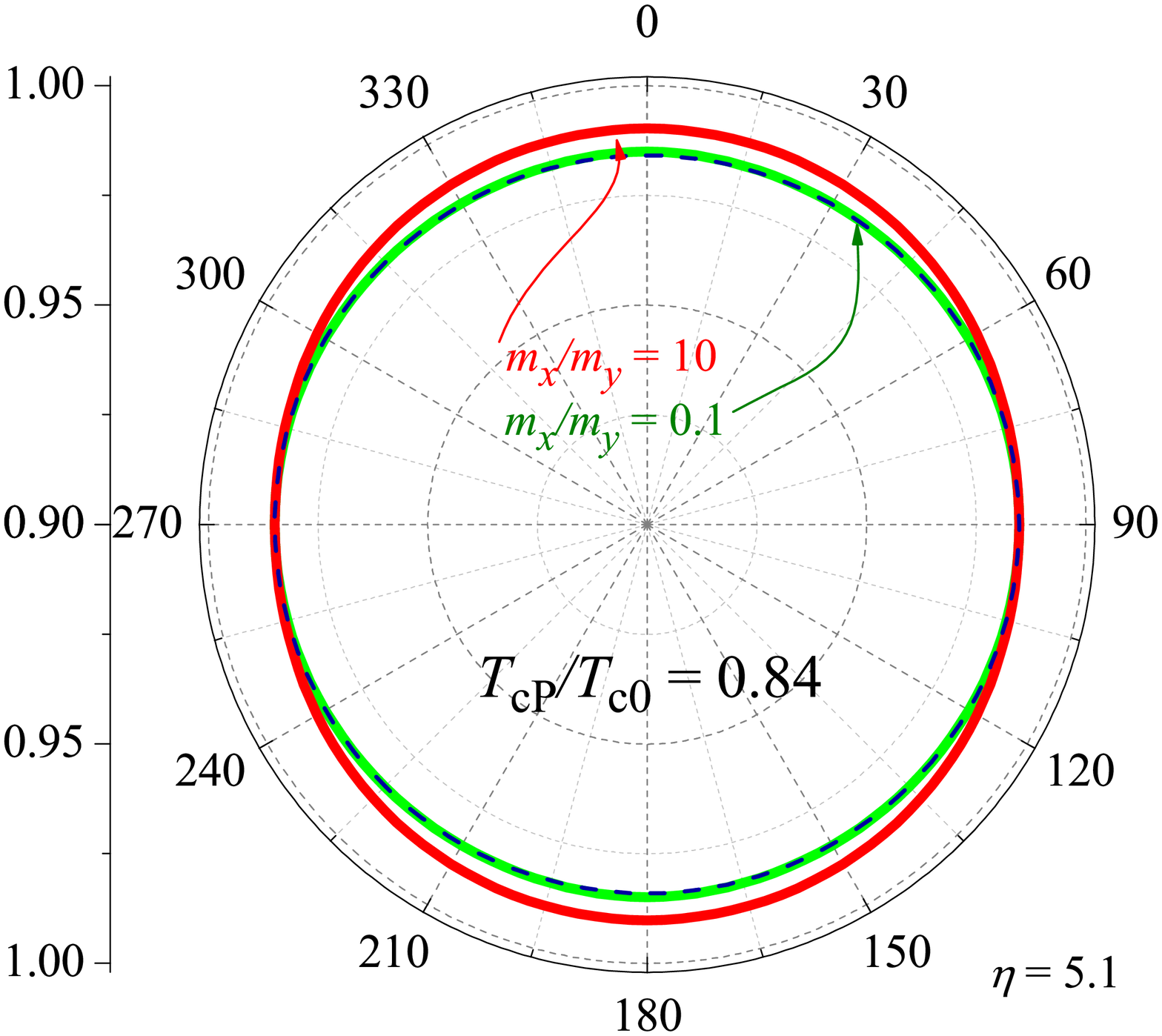}}}
\resizebox{0.5\columnwidth}{!}{\rotatebox{0}{
\includegraphics{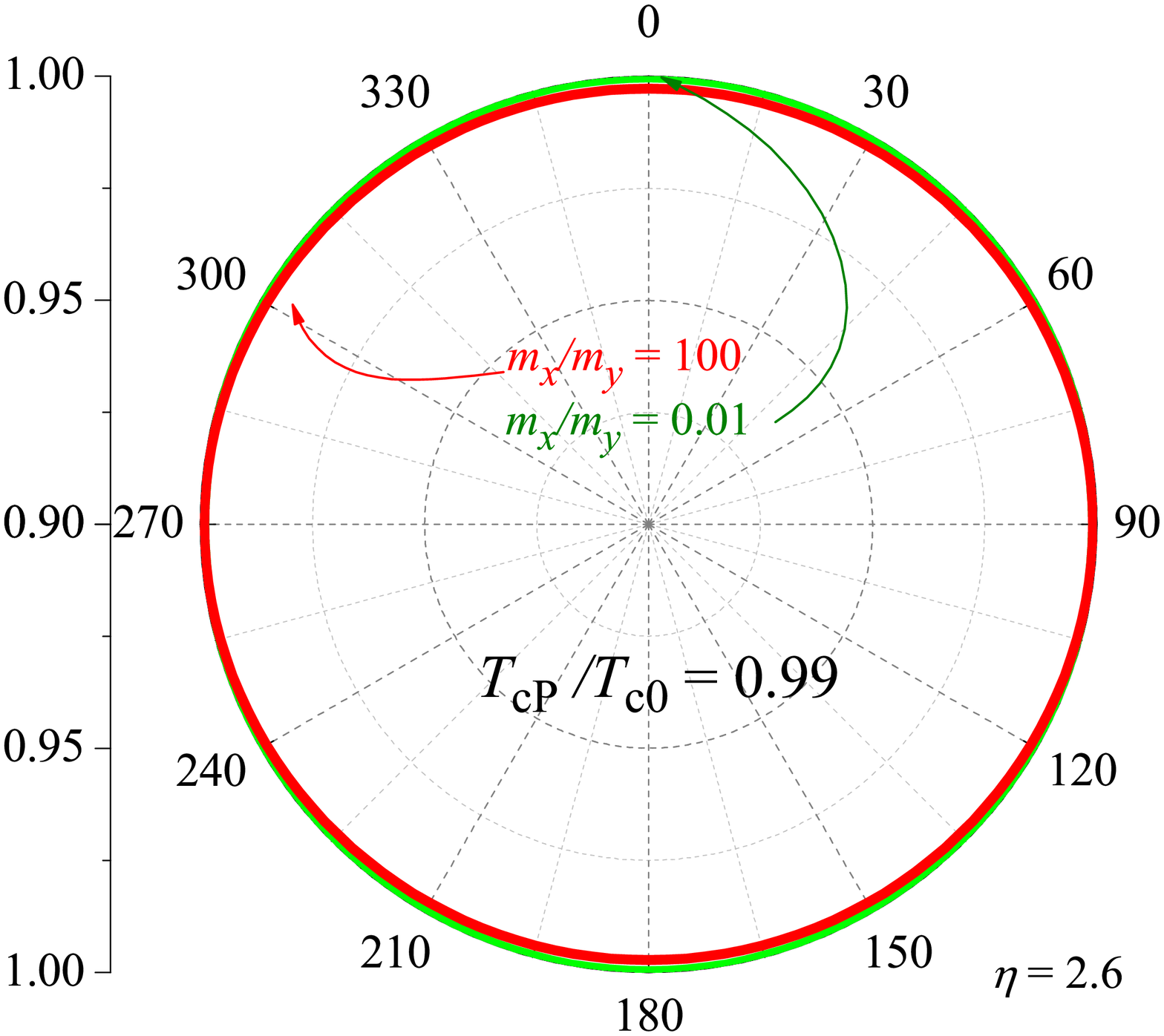}}}
\caption{The same as in Fig. \protect\ref{fig.05} but calculated for $%
\protect\eta =5.1$, which corresponds to $v_{F}=1.5\times 10^{5}~\mathrm{%
m/\sec }$. }
\label{fig.08}
\end{figure*}

\subsubsection{Regime of high magnetic fields $\left( H\gg \frac{T_{c0}}{%
\protect\pi \hbar dv_{F}}\protect\phi _{0}\right) $}

In the absence of the Zeeman effect, when studying the anisotropy of the
upper critical field, such as $T_{c0}\ll \hbar v_{F}Q$, the second harmonics
in the expansion Eq. (\ref{050}) can be neglected, i.e. in Eq. (\ref{080})
we set $\Delta _{\mathbf{\pm }2}=0$ and we get the following equation%
\begin{equation}
\ln \frac{T_{c0}}{T_{c}}=\pi T_{c}\sum\limits_{n}\frac{t^{2}}{\omega _{n}}%
\frac{1}{\omega _{n}^{2}+\hbar ^{2}\left( \frac{\mathbf{v}_{F}.\mathbf{Q}}{2}%
\right) ^{2}}.  \label{092}
\end{equation}%
Performing average over the Fermi surface results in%
\begin{equation}
\ln \frac{T_{c0}}{T_{c}}=\pi T_{c}\sum\limits_{n}\frac{t^{2}}{\omega _{n}^{2}%
}\frac{1}{\sqrt{\omega _{n}^{2}+\left( \hbar v_{F}Q\right) ^{2}/4}}.
\label{093}
\end{equation}%
For extremely large magnitude of the external magnetic field we can
simplify, since $\omega _{n}\sim $ $T_{c0}\ll v_{F}Q$
\begin{equation}
\ln \frac{T_{c0}}{T_{c}}=\frac{t^{2}}{\pi T_{c0}}\sum\limits_{n>0}\frac{1}{%
(n+\frac{1}{2})^{2}}\frac{1}{\hbar v_{F}Q}=\frac{\pi t^{2}}{2T_{c0}}\frac{1}{%
\hbar v_{F}Q}.  \label{094}
\end{equation}%
Therefore, the upper critical field is ($\varkappa _{IV}:$ $H\gg \frac{T_{c0}%
}{\pi \hbar dv_{F}}\phi _{0}$)%
\begin{equation}
\left. H_{c2}^{h=0}\left( \frac{\pi }{2}\right) \right\vert _{\varkappa
_{IV}}=\frac{t^{2}}{2\hbar dv_{F}}\frac{\phi _{0}}{\left(
T_{c0}-T_{c}\right) }.  \label{096}
\end{equation}%
From Eq. \ref{094} it is seen that an increase of the external field far
beyond the value $\frac{T_{c0}}{\pi \hbar dv_{F}}\phi _{0}$ results in a
critical temperature $T_{c}\rightarrow T_{c0}$. Hence at high magnetic
fields the restoration of superconductivity is possible if the destruction
of spin-singlet state of Cooper pairs may be neglected, as was predicted by
Lebed.\cite{lebe01}\cite{lebe07} Therefore, we can infer that within our
model the re-entrant phase of superconductivity is naturally described.
Summarizing the above two sections we plot all considered regimes for the
case of absence of the Zeeman effect in Fig. \ref{fig.03}.


\section{Anisotropy of the upper critical field}

\label{sec4}

In our numerical investigations we restrict ourselves to the following
parameters: the interlayer coupling is $t=2.27$ K, $t/T_{c0}=0.25$, $\Delta
_{0}=2.8kT_{c0}$\cite{mul} and the Fermi velocity $v_{F}=5.0\times 10^{4}~%
\mathrm{m/\sec }$.\cite{iza} Introducing the dimensionless Fermi velocity
parameter, $\eta =\hbar v_{F}\pi d/\phi _{0}\mu _{B}$, this value of $v_{F}$
corresponds to $\eta =1.7$ and $d=1.62$ nm.\cite{bergk} The summation over
the Matsubara frequencies was performed numerically.

Fig. \ref{fig.04} shows the reduced temperature, $T_{cP}/T_{c0}$, dependence
of the magnetic wave vector $\hbar Q_{c2}^{P}v_{F}/k_{B}T_{c0}$ for several
values of the Fermi velocity parameter, when only the paramagnetic effect is
accounted for. Here $Q_{c2}^{P}=\pi dH_{c2}^{P}/\phi _{0}$. The absolute
value of the FFLO modulation wave vector is also given and it grows from
zero for $T<T^{\ast }$. To highlight the contribution of the orbital
correction to the superconducting onset temperature, obtained in the
paramagnetic limit, $\Delta T_{cP}=T_{c}-T_{cP}$, and how it depends on the
magnitude of the external magnetic field applied parallel to the conducting
planes we performed calculations with Eq. (\ref{086}). Fig. \ref{fig.05}
displays the normalized orbital correction, $\Delta T_{cP}/T_{cP}$, as a
function of reduced temperature for several angles $\alpha $\ that the
external field makes from the \textit{x}-axis. The left and middle panels
display the results for the velocity parameter $\eta =1.7$ and $\eta =2.55$,
respectively. The solid lines correspond to the in-plane mass anisotropy $%
m_{x}/m_{y}=100$, while the dashed lines display the results for $%
m_{x}/m_{y}=0.01$. The right panel illustrates the results for $\eta =5.1$, $%
m_{x}/m_{y}=10$ (solid lines), $m_{x}/m_{y}=0.1$ (dashed lines). One can
distinguish the in-plane mass anisotropy from the temperature dependence of
the orbital corrections for angles $\alpha \neq \pm 90%
{{}^\circ}%
$. For example, for $m_{x}/m_{y}=100$ a decrease of temperature from $%
T\lesssim 0.9T_{c0}$, or an increase of the applied magnetic field from $%
H\gtrsim 0.1H_{P0}$, first exhibits a weak influence on $\Delta
T_{cP}/T_{cP} $, but when $T\lesssim 0.65T_{c0}$ ($H\gtrsim 0.5H_{P0}$. Here
$H_{P0}=\Delta _{0}/\mu _{B}$ is the critical magnetic field at $T=0$ in
Pauli limited 2D superconductors) it gradually increases $\left\vert \Delta
T_{cP}\right\vert $, i.e. the orbital suppression of superconductivity
becomes stronger with magnetic field, when orbital pair-breaking is
superimposed on the spin pair breaking mechanism. For $m_{x}/m_{y}=0.01$ an
increase of the applied field results first in a progressive increase of $%
\left\vert \Delta T_{cP}\right\vert $. However, for $T\lesssim 0.65T_{c0}$
we see an opposite bias, namely strengthening of the applied field rapidly
reduces $\left\vert \Delta T_{cP}\right\vert $, \textit{i.e.} the orbital
pair breaking becomes weaker with the external field, and it can almost
vanish for some directions of the field in the very close vicinity of the
tricritical point as seen for dashed curves $\alpha =0$. For $\alpha =90%
{{}^\circ}%
$ the curves describing $m_{x}/m_{y}=100$ mass anisotropy coincide with
those for $m_{x}/m_{y}=0.01$ and both follow the tendency typical for $%
m_{x}/m_{y}=0.01$ mass anisotropy. In Fig. \ref{fig.05} both curves are
given by the thick lines. We can also infer that an increase of the Fermi
velocity weakens this effect of $\left\vert \Delta T_{cP}\right\vert $\
reduction as seen from the middle panel of Fig. \ref{fig.05}. In the FFLO
phase, for $T<T^{\ast }$, or $H>H^{\ast }$, the orbital correction in both
cases of mass anisotropy essentially increases, especially for $%
m_{x}/m_{y}=0.01$ and for some angles can show a non-monotonic behavior. The
further increase of the Fermi velocity can modify the just described
behavior. Indeed, as seen from the right figure the $\alpha =90%
{{}^\circ}%
$ curves follow the tendency typical for $m_{x}/m_{y}=10$ mass anisotropy
and in the FFLO phase they show an upturn.

Opposite tendency in the field direction dependence of the normalized
correction, $\Delta T_{cP}$, for the range of angles $\alpha =0%
{{}^\circ}%
-$ $70%
{{}^\circ}%
$ and for the angles in the close vicinity of $\alpha =90%
{{}^\circ}%
$ in the case of $m_{x}/m_{y}=100$\ should result in a particular anisotropy
of the onset of superconductivity. Figs. \ref{fig.06} and \ref{fig.07} show
the magnetic field angular dependence of the normalized superconducting
transition temperature, $T_{c}\left( \alpha \right) /T_{cP}$, calculated at $%
T_{cP}/T_{c0}\simeq 0.1$, $0.2$, $0.4$, $0.54$, $0.57$, $0.65$, $0.84$ and $%
0.99$ for the velocity parameter $\eta =1.7$ and $\eta =2.55$, respectively.
In the polar plot the direction of each point seen from the origin
corresponds to the magnetic field direction and the distance from the origin
corresponds to the normalized critical temperature. We see that for $%
m_{x}/m_{y}=100$ the reduction of the orbital suppression of
superconductivity at $\alpha =\pm 90%
{{}^\circ}%
$ in the vicinity of the tricritical point is accompanied by a grow of cusps
at these angles\ in the field-angle dependence of $T_{c}\left( \alpha
\right) /T_{cP}$. The cusps appear at $Q\Vert Ox$, \textit{i.e.} magnetic
field is along the light mass direction, as intuitively expected, since it
is more difficult to induce diamagnetic currents with heavier charge
carriers. For $m_{x}/m_{y}=0.01$ the overall orbital corrections are smaller
than that for $m_{x}/m_{y}=100$. This is due to the fact that in the former
case the Fermi surface is smaller and hence the diamagnetic response is
weaker than that in the latter situation. In $\frac{ig.}{{}}$\ref{fig.08} $%
T_{c}\left( \alpha \right) /T_{cP}$ is shown for $\eta =5.1$ and $%
m_{x}/m_{y}=10$ (red lines), $m_{x}/m_{y}=0.1$ (green lines). Formation of
cusps in the vicinity of the tricritical point is also observed, although to
a smaller extent. In Figs. \ref{fig.06}, \ref{fig.07} and \ref{fig.08} the
dashed lines are $T_{c}\left( \alpha \right) /T_{cP}$\ obtained for $%
m_{x}/m_{y}=0.01$($0.1$ in Fig. \ref{fig.08}) when the r.h.s. of Eq. (\ref%
{080}) is neglected, $\Delta _{\mathbf{\pm }2}=0$. In this case the solution
(\ref{086}) simplifies to%
\begin{equation}
T_{c}=T_{cP}\left[ 1-At^{2}a\right]  \label{098}
\end{equation}%
and such solution is valid for $\sqrt{tT_{c0}}\ll \hbar v_{F}Q$, which is
the beginning of the superconductivity re-entrant regime.\cite{lebe01,lebe06}
As the charge carrier mass becomes smaller the superconducting re-entrant
phase begins at a higher magnetic field. Since, according to Eq. (\ref{208})
the second harmonics of the order parameter generates the Lowerence-Doniach
term in the original expression, Eq. (\ref{220}), the dashed lines give a
hint about its contribution to the in-plane anisotropy of the onset of
superconductivity in layered structures with $m_{x}/m_{y}=0.01$ in-plane
mass anisotropy. We see that the difference between the solutions (\ref{086}%
) and (\ref{098}) is negligible for $T_{cP}/T_{c0}\simeq 0.57$. However it
is noticeable already for $T_{cP}/T_{c0}\simeq 0.65$. The upper and lower
knobs are observed when the full original expression is used, and they are
absent for the simplified version, Eq. (\ref{098}). So we can infer that the
observed knobs are due to the Lowerence-Doniach term. Because this term
becomes less important with the field, the knobs are absent for $%
T_{cP}/T_{c0}\simeq 0.57$\ and essentially pronounced for $%
T_{cP}/T_{c0}\simeq 0.85$, when $m_{x}/m_{y}=0.01$. Inversely, for $%
m_{x}/m_{y}=100$ the cusps are profound near the tricritical point,
insignificant for smaller fields, and essentially seen far beyond the
tricritical point in the FFLO\ phase. The cusps are induced by the $t^{2}a$%
-term, which in the conventional phase acquires the following form%
\begin{equation}
at^{2}=\pi T_{cP}\sum\limits_{n}\frac{t^{2}}{\Omega _{n}^{3}}\frac{1}{\sqrt{%
1+\varepsilon \left( \mathbf{Q}\right) /2\Omega _{n}^{2}}}
\end{equation}%
From Fig. (\ref{fig.07}) we can infer that an increase of the Fermi velocity
leads to a narrowing of the cusp width. However such increase of the Fermi
velocity makes the cusps less pronounced.

In the FFLO phase $\hbar v_{F}Q\gtrsim T_{c0}$, and the solution Eq. (\ref%
{098}) can be used for calculations. The top panels of Figs. \ref{fig.06}, %
\ref{fig.07} and \ref{fig.08} illustrate the anisotropy of the
superconducting onset temperature in the FFLO phase. We see that the cusps
induced by the $t^{2}a$-term becomes even more profound with the magnetic
field. Moreover, for $m_{x}/m_{y}=0.01$ a difference between the results
obtained within $\Delta _{\mathbf{\pm }2}\neq 0$ and $\Delta _{\mathbf{\pm }%
2}=0$ appears. For $m_{x}/m_{y}=0.1$ this discrepancy is also present,
although less visible. As was shown and explained in Ref. \cite{vrnkch02}
this deviation this time is due to the resonance between FFLO modulation
wave vector and the interlayer coupling modulated by the vector potential.
Thus, in addition to the overall anisotropy induced by the FFLO modulation
and studied in Ref. \cite{vrnkch01}, additional cusps develop for certain
directions of the applied field, when the resonance conditions are realized.
To describe resonances we have to account for the second harmonics, $\Delta
_{\mathbf{\pm }2}$, and then
\begin{multline}
S^{\pm }\left( \mathbf{Q}\right) \equiv \frac{\left( a+b_{\pm }\right)
t^{2}+\delta _{\pm }}{2}  \notag \\
+\frac{t^{2}}{2}\sqrt{\left[ a-b_{\pm }-\delta _{\pm }/t^{2}\right]
^{2}+4c_{\pm }^{2}}.
\end{multline}%
In general, in the vicinity of the tricritical point when comparing the
in-plane anisotropy of $T_{c}\left( \alpha \right) $ for the conventional
phase with that in the FFLO modulated phase, $T<T^{\ast }$ or $H>H^{\ast }$,%
\cite{vrnkch01} it is obviously seen a significant discrepancy. On both
sides of the tricritical point, $T^{\ast }$, the contribution of the $t^{2}a$%
-term is essential and the observed difference is purely induced by the
appearance of the FFLO modulation wave vector.

The anisotropy of the onset of superconductivity obtained within our model
for $t\ll \hbar v_{F}Q$ and $m_{x}/m_{y}=100$\ qualitatively similar to that
observed in the experiment with (TMTSF)$_{2}$ClO$_{4}$.\cite{yone01} For $%
H<H^{\ast }$ our theoretical calculations show that in $T_{c}\left( \alpha
\right) /T_{cP}$ cusps develop along the light masses. The same cusps and
along the this direction are visible for $H=20$ kOe and $H=25$ kOe in the
experimental data for $T_{c}\left( \alpha \right) /T_{cP}$. Our calculations
show that for $H>H^{\ast }$ small dips appear from both sides of each cusp.
Similar picture is observed in the experiment for $H>30$ kOe.

If we compare the field-direction dependence of the superconducting onset
temperature for $T_{cP}/T_{c0}\lesssim 0.85$, valid for $\hbar v_{F}Q$ $\gg
t $, with that in the last panels of Figs. \ref{fig.06}, \ref{fig.07} and %
\ref{fig.08}, where the result of the Ginzburg-Landau regime Eq. (\ref{020}%
), valid for $\hbar v_{F}Q$ $\ll t$, is shown at $T_{cP}/T_{c0}\simeq 0.99$
we see an essential distinction. In the vicinity of $T_{c0}$ the anisotropy
of the onset of superconductivity shows a typical picture for the
anisotropic Ginzburg-Landau model. $T_{c}\left( \alpha \right) $ is maximum
for $\mathbf{H}\bot Ox$ near $T_{c0}$ and as seen from Fig. \ref{fig.06}
also in the vicinity of $T^{\ast }$.

\section{Conclusions}

\label{sec6}

In this work we have derived the extended Lawrence-Doniach model, which
allows one to study superconductivity of layered materials at high magnetic
fields. Within this model we have analyzed the field-amplitude and the
field-direction dependence of the onset of superconductivity in layered
conductors. Our theoretical analysis gives rise to the following assertion.
There are four regimes, which we discriminate according to the distinctive
features of the anisotropy of the onset of superconductivity and the
temperature dependence of the upper critical field. (\textit{i}) In the
Ginzburg-Landau regime, when $H\ll \frac{t}{\pi \hbar dv_{F}}\phi _{0}$, $%
\left. H_{c2}\right\vert _{GL}\sim \left( T_{cP}-T_{c}\right) $, the
anisotropy is well described within the continuous GL model. (\textit{ii})
In the Lowerence-Doniach regime, within $t\phi _{0}/\pi \hbar dv_{F}\ll H\ll
\sqrt{tT_{c0}}\phi _{0}/\pi \hbar dv_{F}$, $\left. H_{c2}\right\vert
_{LD}\sim 1/\sqrt{\left( T_{c}-T_{ct}\right) }$, the anisotropy is mostly
determined by the term proportional to $t^{4}/\left( \hbar v_{F}Q\right)
^{2} $, which induces knobs in the direction along the light masses in the
field-angle dependence of $T_{c}\left( \alpha \right) $. (\textit{iii}) For $%
\frac{\phi _{0}}{\pi \hbar dv_{F}}\sqrt{tT_{c0}}\ll H\ll \frac{\phi _{0}}{%
\pi \hbar dv_{F}}T_{c0}$, $\left. H_{c2}\right\vert _{RS}=\sqrt{\left(
T_{c}-T_{ct}\right) }$, the anisotropy is governed by the $t^{2}a$-term,
which is responsible for the re-entrant of superconductivity. (\textit{iv})
the FFLO phase, $H>H^{\ast }$, the anisotropy is settled by the interplay
between the modulation and magnetic field wave vectors. The third regime can
be deep in the four one so the discussed cusps can be invisible in the
conventional phase. The paramagnetic effect is crucial for the description
of the upper critical field both above and below the tricritical point. If
the paramagnetic effect is negligible than the extended Lowerence-Doniach
model restores the re-entrant behavior with magnetic field originally
obtained by Lebed.\cite{lebe01,lebe06}

Near $T_{c0}$ the anisotropy of the onset of superconductivity shows the
smooth variation of $T_{c}\left( \alpha \right) $. When reducing the
temperature, above the tricritical point small cusps appear. We may expect
that small cusps observed in the field-direction dependence of $T_{c}\left(
\alpha \right) /T_{cP}$\ in the experiment with (TMTSF)$_{2}$ClO$_{4}$ near
the Pauli limiting field, $H_{P0}=26$ kOe\cite{yone01}\ could have the
re-entrant phase origin and are well described by the extended
Lawrence-Doniach model. A technique that control the anisotropy of the upper
critical field can provide an invaluable tool for investigating the physical
origin of the experimentally observed upturn of the upper critical field in
the low temperature regime.

\begin{acknowledgments}
We acknowledge the support by the European Community under a Marie Curie IEF
Action (Grant Agreement No. PIEF-GA-2009-235486-ScQSR) and European IRSES
program SIMTECH.
\end{acknowledgments}

\appendix


\section{Derivation of the expression for A.}

\label{sec:appendix_A}

Substitution of Eq. (\ref{015a}) in Eq. (\ref{014}) results in
\begin{multline}
\Delta \left( \mathbf{r},k_{z}\right) \ln \frac{T_{c}}{T_{cP}}=\Delta \left(
\mathbf{r},k_{z}\right) \\
\times \left[ F\left( \frac{h}{\pi T_{c}}\right) -F\left( \frac{h}{\pi T_{cP}%
}\right) \right] +\widehat{\Pi }_{\mathrm{MLD}}\Delta \left( \mathbf{r}%
,k_{z}\right) ,
\end{multline}%
where we defined a function%
\begin{equation}
F\left( \frac{h}{\pi T}\right) \equiv \pi T\sum\limits_{n}\left[ \frac{1}{%
\omega _{n}\left( T\right) }-\frac{1}{\Omega _{n}\left( T\right) }\right] .
\end{equation}%
When expanding in series, taking into account that $\left(
T_{c}-T_{cP}\right) /T_{c}\ll 1$ we obtain
\begin{multline}
\Delta \left( \mathbf{r},k_{z}\right) \frac{T_{c}-T_{cP}}{T_{c}}=\Delta
\left( \mathbf{r},k_{z}\right) \frac{h}{\pi T_{cP}}\frac{T_{c}-T_{cP}}{T_{c}}
\\
\times \frac{\partial }{\partial \left( \frac{h}{\pi T}\right) }\left.
F\left( \frac{h}{\pi T}\right) \right\vert _{T=T_{cP}}+\widehat{\Pi }_{%
\mathrm{MLD}}\Delta \left( \mathbf{r},k_{z}\right) ,
\end{multline}%
and hence%
\begin{equation}
\Delta \left( \mathbf{r},k_{z}\right) \frac{T_{c}-T_{cP}}{AT_{c}}=\widehat{%
\Pi }_{\mathrm{MLD}}\Delta \left( \mathbf{r},k_{z}\right) ,
\end{equation}%
where we introduced the following notations%
\begin{equation}
P=\frac{T_{c}-T_{cP}}{AT_{c}},  \label{309}
\end{equation}%
and $A$\ is given by%
\begin{equation}
A^{-1}=1-\frac{h}{\pi T}\frac{\partial }{\partial \left( \frac{h}{\pi T}%
\right) }\left. F\left( \frac{h}{\pi T}\right) \right\vert _{T=T_{cP}}.
\label{310}
\end{equation}


\section{Derivation of Eqs. (\protect\ref{080}-\protect\ref{082})}

\label{sec:appendix_B}

Solution of the system of coupled equations (\ref{060} - \ref{063}) can be
found as follows. From Eq. (\ref{063}) we find%
\begin{equation}
f_{\pm 3}=\mp \frac{\widetilde{t}f_{\pm 2}}{L_{n}\left( \pm 3\mathbf{Q}%
\right) }
\end{equation}%
and substituting it into Eq. (\ref{062}) gives%
\begin{equation}
\left[ L_{n}\left( \pm 2\mathbf{Q}\right) +\frac{t^{2}}{L_{n}\left( \pm 3%
\mathbf{Q}\right) }\right] f_{\pm 2}\pm \widetilde{t}f_{\pm 1}=\Delta _{%
\mathbf{\pm }2}.  \label{064}
\end{equation}%
Then substitution of $f_{\pm 1}$, obtained from Eq. (\ref{061}),
\begin{equation}
f_{\pm 1}=\mp \frac{\widetilde{t}f_{0}}{L_{n}\left( \pm \mathbf{Q}\right) }%
\mathbf{\pm }\frac{\widetilde{t}f_{\pm 2}}{L_{n}\left( \pm \mathbf{Q}\right)
},  \label{066}
\end{equation}%
when taking into account that within the required approximation $%
f_{0}\approx \Delta _{0}/L_{n}\left( \mathbf{q}\right) $, produces the
equation for the second harmonic of the pair amplitude, $f_{\pm 2}$,
\begin{multline}
\left[ L_{n}\left( \pm 2\mathbf{Q}\right) +\frac{\widetilde{t}^{2}}{%
L_{n}\left( \pm 3\mathbf{Q}\right) }+\frac{\widetilde{t}^{2}}{L_{n}\left(
\pm \mathbf{Q}\right) }\right] f_{\pm 2} \\
-\frac{\widetilde{t}^{2}\Delta _{0}}{L_{n}\left( 0\right) L_{n}\left( \pm
\mathbf{Q}\right) }=\Delta _{\mathbf{\pm }2}.  \label{069}
\end{multline}%
Substitution of $f_{\pm 1}$ from Eq. (\ref{066}) and $f_{\pm 2}\approx
\Delta _{\pm 2}/L_{n}\left( \mathbf{q}\pm 2\mathbf{Q}\right) $, obtained
within the required approximation from Eq. (\ref{069}), into Eq. (\ref{060})
results in the following equation for $f_{0}$
\begin{multline}
\left[ L_{n}\left( 0\right) +\frac{\widetilde{t}^{2}}{L_{n}\left( +\mathbf{Q}%
\right) }+\frac{\widetilde{t}^{2}}{L_{n}\left( -\mathbf{Q}\right) }\right]
f_{0} \\
-\sum\limits_{\pm }\frac{\widetilde{t}^{2}\Delta _{\mathbf{\pm }2}}{%
L_{n}\left( \pm \mathbf{Q}\right) L_{n}\left( \pm 2\mathbf{Q}\right) }%
=\Delta _{0}.  \label{072}
\end{multline}%
Since we adopt a second-order approximation in the small parameter $t/T_{c0}$
Eqs.(\ref{069} - \ref{072}) acquire the following form
\begin{multline}
f_{0}=\Delta _{0}\left[ \frac{1}{L_{n}\left( 0\right) }-\frac{\widetilde{t}%
^{2}}{L_{n}^{2}\left( 0\right) L_{n}\left( +\mathbf{Q}\right) }\right. \\
-\left. \frac{\widetilde{t}^{2}}{L_{n}^{2}\left( 0\right) L_{n}\left( -%
\mathbf{Q}\right) }\right] \\
+\sum\limits_{\pm }\frac{\widetilde{t}^{2}\Delta _{\mathbf{\pm }2}}{%
L_{n}\left( 0\right) L_{n}\left( \pm \mathbf{Q}\right) L_{n}\left( \pm 2%
\mathbf{Q}\right) }.  \label{075}
\end{multline}%
\begin{multline}
f_{\pm 2}=\Delta _{\mathbf{\pm }2}\left[ \frac{1}{L_{n}\left( \pm 2\mathbf{Q}%
\right) }-\frac{\widetilde{t}^{2}}{L_{n}^{2}\left( \pm 2\mathbf{Q}\right)
L_{n}\left( \pm 3\mathbf{Q}\right) }\right. \\
-\left. \frac{\widetilde{t}^{2}}{L_{n}^{2}\left( \pm 2\mathbf{Q}\right)
L_{n}\left( \pm \mathbf{Q}\right) }\right] \\
+\frac{\widetilde{t}^{2}\Delta _{0}}{L_{n}\left( 0\right) L_{n}\left( \pm
\mathbf{Q}\right) L_{n}\left( \pm 2\mathbf{Q}\right) }.  \label{079}
\end{multline}%
Submitting the obtained expressions for $f_{0}$ and $f_{\pm 2}$ back into
the self-consistency relation Eq. (\ref{008}) results in Eqs. (\ref{080}-\ref%
{082}).


\section{Derivation of Eq. (\protect\ref{089})}

\label{sec:appendix_C}

If $\left( \hbar v_{F}Q\right) \ll T_{c0}$, or $H\ll \frac{\phi _{0}}{\pi
\hbar dv_{F}}T_{c0}$, then $P+t^{2}b_{\pm }\ll \delta _{\pm }$ and we find
from Eq.(\ref{081}) that
\begin{equation}
\Delta _{\mathbf{\pm }2}\approx \frac{t^{2}c_{\pm }}{\delta _{\pm }}\Delta
_{0},  \label{201}
\end{equation}%
with [see Eqs. (\ref{083}) and (\ref{084})]%
\begin{align}
\delta _{\pm }& =\pi T_{cP}\sum\limits_{n}\frac{1}{\Omega _{n}}\left[ 1-%
\frac{1}{\sqrt{1+g^{2}}}\right] , \\
c_{\pm }& =\pi T_{cP}\sum\limits_{n}\frac{1}{\Omega _{n}^{3}}\left[ \frac{1}{%
\sqrt{1+g^{2}}}-\frac{1}{\sqrt{4+g^{2}}}\right] ,
\end{align}%
where $g\equiv \hbar v_{F}Q/\Omega _{n}$. Expansion of these expressions
with respect to $g\ll 1$ gives%
\begin{align}
\delta _{\pm }& \approx \pi T_{cP}\sum\limits_{n}\frac{\left( \hbar
v_{F}Q\right) ^{2}}{2\Omega _{n}^{3}}, \\
c_{\pm }& \approx \pi T_{cP}\sum\limits_{n}\frac{1}{2\Omega _{n}^{3}}\left[
1-\frac{7\left( \hbar v_{F}Q\right) ^{2}}{8\Omega _{n}^{2}}\right] ,
\label{208}
\end{align}%
and from Eq. (\ref{081}) we find that $\Delta _{\mathbf{\pm }2}$ reads as
\begin{equation}
\Delta _{\mathbf{\pm }2}\approx \frac{t^{2}}{\left( \hbar v_{F}Q\right) ^{2}}%
\Delta _{0}.  \label{210}
\end{equation}%
Substitution of $\Delta _{\mathbf{\pm }2}$ back into Eq. (\ref{080}) leads
to the following equation, determining temperature $T_{c}$ of the onset of
the superconducting state, when the orbital effects of the applied magnetic
field are accounted for within the second-order approximation in parameter $%
t/T_{c0}$,
\begin{equation}
P+t^{2}a=\frac{t^{4}}{\left( \hbar v_{F}Q\right) ^{2}}\sum\limits_{\pm
}c_{\pm },
\end{equation}%
where $a=2\pi T_{cP}\sum\nolimits_{n}1/\Omega _{n}^{3}\sqrt{4+g^{2}}$.
Making use of the expansion of $a$ into a series%
\begin{equation}
a\approx \pi T_{cP}\sum\limits_{n}\frac{1}{\Omega _{n}^{3}}\left[ 1-\frac{1}{%
8}\frac{\left( \hbar v_{F}Q\right) ^{2}}{\Omega _{n}^{2}}\right] ,
\end{equation}%
we obtain equation for $T_{c}$
\begin{equation}
P=-\pi T_{cP}\sum\limits_{n}\frac{t^{2}}{\Omega _{n}^{3}}\left[ 1-\frac{1}{8}%
\frac{\left( \hbar v_{F}Q\right) ^{2}}{\Omega _{n}^{2}}-\frac{t^{2}}{\left(
\hbar v_{F}Q\right) ^{2}}\right] .  \label{220}
\end{equation}%
After introducing $T_{ct}$, as it is done in Ref. (\ref{035}), which
accounts for the coupling between adjacent layers, finally we obtain Eq. (%
\ref{089}).


\end{document}